\documentclass[aps,pre,twocolumn]{revtex4-1}
\usepackage{graphicx}
\usepackage{bm}
\usepackage{amsmath}
\usepackage{esvect}
\usepackage[utf8]{inputenc} 
\usepackage[T1]{fontenc}
\usepackage{textcomp}

\DeclareUnicodeCharacter{2008}{\ensuremath{-}}

\begin{document}



\title{ Bi-Hamiltonian in Semiflexible Polymers built upon Overdamping Process}

\author{Heeyuen Koh}
\email{heeyuen.koh@gmail.com}
\affiliation{Soft Foundry Institute, Seoul National University, 1 Gwanak-ro, Gwanak-gu, Seoul, 08826, South Korea}
\author{Shigeo Maruyama}
\affiliation{Mechanical Engineering Dept., The University of Tokyo, Hongo 3-7-1, Bunkyoku, Tokyo, Japan}

\date{\today}

\begin{abstract}
 Quantifying the interaction between a system of interest and its ambient conditions, the memory effect links the states of two distinct Hamiltonians: one for the target system and one for the environment.  In this paper, we propose a diffusion process derived from the Smoluchowski equation that can describe the evolution process described by the memory effect integration in a non-Markovian regime.  The Smoluchowski picture, within the framework of stochastic thermodynamics, justifies a diffusion process incorporated into the equations of motion, and the result of the derivation enables a coarse-grained molecular dynamics simulation with the modified equation of motion to reproduce attenuation from collisions between single walled carbon nanotubes (SWCNTs) under far from equilibrium conditions. The results of the numerical experiments on the collision confirm that heat diffusion compensates for the correlated momentum arising from the memory effect of a strongly coupled system with a bi-Hamiltonian structure in both equilibrium and far from equilibrium states.\end{abstract}

\maketitle
\section{Introduction}
Memory effect integral\cite{Zwanzig1973}  that defines the intercorrelation between two distinguished Hamiltonians for the target system and the heat bath, derives the fluctuation dissipation theorem under a Markovian system. The bi-Hamiltonian structure\cite{Magri1978, Olver1987} between the target system and the heat bath, whose dynamics is affected by two independent Hamiltonians at the same time,  derived as the memory effect\cite{Zwanzig1973} under Mori-Zwanzig formalism, its dynamics is valid in far-from-equilibrium conditions or non Markovian regimes where the classical fluctuation-dissipation theorem ceases to hold. Therefore, it has been regarded as describing the evolution of dynamics along the nonlinearity underlying chaotic behavior in discrete systems\cite{Gallas1993,Peters2022,DelPino2024}, polymeric systems\cite{Panja_2010,Klippenstein_2021, Schmid_2023}, and recent theoretical developments underlying the regulation of open quantum systems\cite{Addis2016,Wang2017,Wang2023}. 
 
  Inspired by the energetics of the Langevin equation following Sekimoto's frameworks\cite{Sekimoto1998} and the Hamiltonian of mean force\cite{Kirkwood1935,Campisi2009a,Seifert2012}, the evolution process from the fluctuation of the environment, especially for small enough systems that are easily affected by such interaction, has been the main concern of stochastic thermodynamics. The measurability of the Hamiltonian of mean force\cite{Ding2023,Talkner2016} has raised arguments\cite{Talkner2016,Strasberg2017,Strasberg2020,Talkner2020} whose quantification parallels the complexities associated with the memory effect integration. Meanwhile, such expression involves distinguishing the fluctuation dynamics of the molecular system, which is thoroughly investigated to quantify the thermodynamic aspects coupled to its evolution in the master equation\cite{Ding2023,Busiello2019,Kawai2016} which governs the irregularity of entropy\cite{Strasberg2017} or nonlinear characteristics at small scales\cite{Ciliberto2017}.

The specification of fluctuation dynamics, like the recent approach on thermodynamics uncertainty relations\cite{Gingrich2016,Horowitz2020,period,Gao2024,Ptaszyski2024}, enriches current investigations on identifying the attributes from nonequilibrium states \cite{Sekimoto1998,Seifert2012,Hill2001,Talkner2016,Jarzynski2011,Jarzynski2017,Celani2012,Broeck2015,Ding2022_}. Thermodynamics speed limit\cite{shiraishi2018speed,lee2022speed} seems to be along the context to incorporate the bi-Hamiltonian structure and its result in memory effect integration into current paradigm of stochastic thermodynamics. Nonetheless, a strategic initial move might involve capturing the memory effect in an easily quantifiable system, one that displays non-Markovian dynamics when subjected to nonequilibrium or strong coupling conditions, thus forging a distinct relationship between the memory effect from the bi-Hamiltonian structure and a refined framework in stochastic thermodynamics.  

   One example of a strongly coupled condition that explicitly demonstrates the memory effect is mode coupling between two independent Hamiltonian systems, as in rotational-translational coupling, which is directly linked to a well-known nonlinear dynamical phenomenon, the Dzhanibekov effect, or the tennis-racket theorem\cite{delaTorre2024,Ostanin2023,Lun-Fu2022,Ashbaugh1991}. Similar coupling has been observed in water molecules in several experiments in the 1970s\cite{Berne1976,Davies1978} and in recent work by Jin and Voth\cite{Jin2023}, whose significance is demonstrated by the correction between two distinct types of motion in coarse grained molecular dynamics (CGMD) simulation of water molecules, thereby improving the precision on diffusivity. 
      
   Koh et al. have shown that translational and rotational coupling of CG particles is described as the memory-effect integration using the Mori-Zwanzig formalism\cite{Koh2021}.  The derivation also offers the empirically optimized diffusion process with double partial derivative incorporated with the equation of motion that replaces the memory effect integration during the CGMD simulation of nonlinear bending motion of single walled carbon nanotubes(SWCNT), and the result yields nearly perfectly synchronized dynamics as observed in the atomic-scale simulation\cite{Koh2021,Koh2021_}.  
   
   The accuracy observed in simulations of water molecular systems\cite{Jin2023} and semi-flexible polymers\cite{Koh2021,Koh2023} from modifications that address mode-coupling anomalies or memory effects merits further investigation to seek a more precise understanding of the strongly coupled condition from bi-Hamiltonian system and the replacement of the memory effect with the diffusion process, which redistributes the correlated energy of the memory effect integration into each Hamiltonian system. When it is involved with various conditions like a nanoscale system in sub-kelvin environment\cite{Vigneau2022,Samanta2023}, a more rigorous theoretical approach on quantification of the memory effect and related thermodynamic framework will provide promising results for further applications on nanomechanical qubits\cite{Arrangoiz-Arriola2019,Pistolesi2021,Wallucks2020}.  

In this paper, the stochastic thermodynamics framework\cite{Jarzynski2017} is adapted to clarify cross correlation from the memory effect\cite{Koh2021}, which is empirically confirmed to be a diffusion process\cite{Koh2021}. The derivation using Smoluchowski equation also justifies the inclusion of the heat-diffusion term in the equation of motion for the CGMD simulation, as confirmed by the SWCNT simulation in far from equilibrium. In section II. Theoretical modeling is derived based on the framework by Jarzynski\cite{Jarzynski2017}. The data provided by molecular dynamics and coarse grained molecular dynamics is explained in Section III. Simulation. The meaning of derivation, which is confirmed from the Simulation, is described in IV. Discussion. The summary and further works are included in V. Conclusion.

\section{Theoretical modeling}
\subsection{Dzanibekov effect in coarse-grained particle}

Let a quasi-one-dimensional system, such as a SWCNT, oscillate laterally as shown in Fig. 1A in a vacuum chamber with the adiabatic condition and two fixations at the boundaries. When the tube is simplified as a series of groups of atoms that are connected with the harmonic potential energy functions for bond length $\Phi_{\ell}$ and the angle between CG particles, $\Phi_{\theta}$, the Hamiltonian of the CG system becomes

\begin{eqnarray}
H = H_{\ell 0}+H_{\theta 0},\label{eq:eq1} \\
H_{\ell 0} = \frac{1}{2m} \mathbf{P}^T_{\ell}\mathbf{P}_{\ell} + \Phi_{\ell},\label{eq:eq2}\\
H_{\theta 0} = \frac{1}{2I} \mathbf{P}_{\theta}^T\mathbf{P}_{\theta} + \Phi_{\theta}, \label{eq:eq3}
\end{eqnarray}

with the set of momentum for the bond length $ \mathbf{P}_{\ell}=\left[\mathbf{P}^{1}_{\ell},..,\mathbf{P}^{N}_{\ell} \right]^T$, and the angle $\mathbf{P}_{\theta}=\left[\mathbf{P}^{1}_{\theta},..,\mathbf{P}^{N}_{\theta} \right]^T$. $T$ means transpose, and $N$ is the total number of CG particles. $m$ is the mass of the particle, and $I$ is the inertia of the particle. The component in blue shade in the tube has the deformation along each axis as shown in Fig. 1A.

 \begin{figure}
 \includegraphics[scale=0.6,trim={0cm 10cm 20cm 0cm},clip]{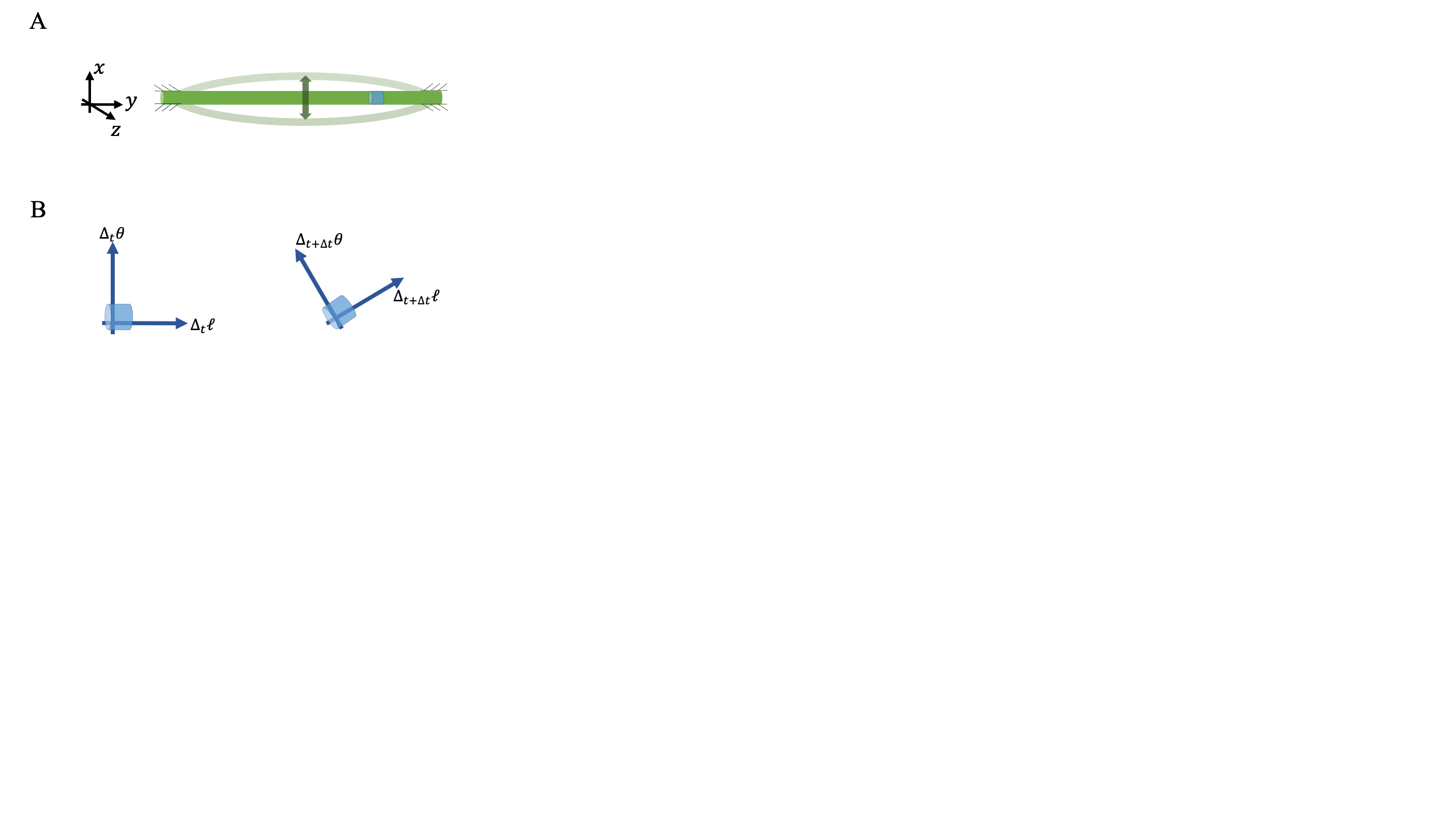}
 \caption{A. Schematic figure of suspended SWCNT and a part of the structure that is equivalent to a CG particle in blue shade. B. Evolution of the deformation along the bond length and angle direction at $t$ and $t+\Delta t$. The coordinate system used to measure deformation evolves with the macroscopic motion of the tube, therefore, the momentum defined along the coordinate system affected by Dzhanibekov effect. $\Delta \ell$ and $\Delta \theta$ are the deformation along the tube length and angle. $\Delta r$ is the radial deformation, which is omitted. }
 \end{figure}
 
Since the coordinate system of a bead in a coarse-grained system is defined for the translation and the rotational change at each time step during the evolution process, as shown in Figure 1B, the momentum that belongs either of $\theta$ or $\ell$ is slightly altered by which unit axes at $t$ and $t+\Delta t$ are used to count the motion. For instance, the bond length-wise deformation at $t$ and that from the angle at $t+\Delta t$ become non-orthogonal to each other, so that the part of the momentum predefined at $t$ is interwoven into another species of the momentum that is specified with the unit axis defined at $t+\Delta t$. In this manner, the trajectory that forms the momentum of two distinct Hamiltonians is constantly affected by each other, as the Dzhanibekov effect has demonstrated with a rotation of the spinning axis of a wingnut in the spaceship\cite{delaTorre2024,Ostanin2023,Lun-Fu2022, Ashbaugh1991}. 

The evolution of the momentum along $\theta$ and $\ell$, which are intercorrelated with each other by the Dzhanibekov effect, is supposed to regulate the oscillation of the suspended SWCNT, ensuring it remains in a state of equilibrium if the given condition of the SWCNT is in the adiabatic environment. Yet, the CG system that is affected by the Dzhanibekov effect during its evolution process is not strictly in the canonical state of a harmonic Hamiltonian for a weak perturbation.  The relation of the unit vectors for each momentum $\mathbf{P}_{\ell}$ and $\mathbf{P}_{\theta}$, which are defined at the different cartesian coordinate at $t$ and $t+\Delta t$ can be written as 

\begin{align}
\hat{\bf e}_{\theta}(t+\Delta t) = a \hat{\bf e}_{\theta}(t)+ a' \hat{\bf e}_{\ell}(t),\label{eq:eq4} \\
\hat{\bf e}_{\ell}(t+\Delta t) = b{\bf e}_{\ell}(t)+ b' \hat{\bf e}_{\theta}(t). \label{eq:eq5}
\end{align}

$a$,$b$,$a'$ and $b'$ are arbtrary given paratmers with the assumption $a >> a'$ and $b >> b'$. $\hat{\bf e}_{\theta}$ and $\hat{\bf e}_{\ell}$ are the unit vectors that are defined along angle and bond length, respectively.

Each Hamiltonian with the momentum defined with Eq.(\ref{eq:eq4}) and Eq.(\ref{eq:eq5}) experiences a slight offset from the trajectory designated by the conservative force from the potential energy landscape. Yet, each Hamiltonian should share the perturbed kinetic energy under the total energy conservation, which is strictly guaranteed from the original system before the coarse graining process. The condition of the perturbation that minimizes the entropy production from coarse graining process could be in the form of the following:  

\begin{align}
H_{\ell} = H_{\ell 0} \pm \Delta H_D\label{eq:eq_perturb1}\\
H_{\theta} = H_{\theta 0} \mp \Delta H_D. \label{eq:eq_perturb2}
\end{align}

Intuitively, the consequence of the perturbation from the Dzhanibekov effect, $\Delta H_D$, to an atomic-scale system should be evanescent events as part of the omitted damping process from the loss of the degrees of freedom when the system is reduced to the CG model. This omitted damping process is intended to maintain the oscillation of the CG system defined with Eq.(\ref{eq:eq1}) with the consistency as measured in atomic scale simulations. 

 The perturbed terms are regarded as the bi-Hamiltonian structure arising from the evolution of two Hamiltonian systems. Therefore, the adjustments derived by the perturbation from Eq.(\ref{eq:eq4}) and Eq.(\ref{eq:eq5}) are expressed in the form of memory effect integration, more precisely, the cross-correlated states in the frequency domain. The Mori-Zwanzig formalism for the cross correlated states and memory effect integration is well shown in the previous study\cite{Koh2021}. The diffusion process, empirically shown to be equivalent to the memory effect integration, is derived within the stochastic thermodynamics framework in the following subsection.

\subsection{Stochastic thermodynamics for cross correlated states }

Due to the adiabatic environment in the vacuum chamber, there is a very limited number of particles, which is far below being regarded as the thermodynamic limit of an infinitely large system in the heat bath or the definition of the partition function. The probability density function from a finite bath \cite{Campisi2009a} starts from the very beginning of its definition, and the partition function of the target system becomes dependent on the heat capacity, $C$, as below:

\begin{align}
\rho(z,\lambda) = \frac{\Omega_B (E_{tot}-H(z,\lambda))}{\Omega_{tot}(E_{tot})}, \label{eq:eq6}\\
\lim_{C \rightarrow \infty} \rho_{C}(z;T,\lambda) = \frac{e^{-H/T}}{Z(T,\lambda)}. \label{eq:eq7}
\end{align}

 Here, $z = \left( \bf{q}, \bf{p}\right)$ is the set of state variables for the system composed of the ideal gas, and $\lambda$ is a controllable parameter. The density of the state of the bath and the total system are $\Omega_{B}(E)$ and $\Omega_{tot}(E)$, respectively. The form of the probability density function, $\rho$ in Eq. (\ref{eq:eq6}) becomes as Eq. (\ref{eq:eq7}) when $C \rightarrow \infty$ as $dn \rightarrow \infty$ where the degree of freedom of the system is $d$ and the number of particles is $n$. 
 
A more specific condition for $  C\rightarrow \infty$ in harmonic oscillator represented with Eq.(\ref{eq:eq1})$\sim$Eq(\ref{eq:eq3}) is derived in Appendix A. The result of the derivation validates the definition of probability density function under the well-known condition $dn \rightarrow \infty$, but it also shows that the finite amount of $dn$ from the number of CG particles around $\mathcal{O}(10^2) \sim \mathcal{O}(10^3)$ can afford the condition close to $C \rightarrow \infty$ that overcomes the limit of the number of particles in our system for SWCNT. 

Then, the ensemble of the system of interest derived from the stochastic thermodynamics framework suggested by Jarzynski\cite{Jarzynski2017} becomes

\begin{align}\label{eqn:ensemble}
\rho(t)  = \frac{1}{\mathcal{Z_{\lambda}}}e^{-\beta (h_s(x_j;\lambda)+\phi(x_j)) },
\end{align}

with 

\begin{align}
\label{eqn:ensemble}
\phi(x_j) = \phi(x_j:N,P,T), \nonumber \\ 
=-\beta^{-1}\frac{ \int dy_j exp[\ -\beta ( H_{\mathcal{E}}(y_j) +h_{SE}  + PV_{\mathcal{E}}) ]\ }{\int dy exp[\ -\beta( H_{\mathcal{E}}(y_j) + PV_{\mathcal{E}} )]\ }, 
\end{align}

and  

\begin{align}\label{eqn:e_result}
\mathcal{Z}_{\lambda} (N,P,T)=\int dx_j e^{-\beta (h_s+\phi(x_j))}.
\end{align}

  $x_j = \left( \bf{Q}_j, \bf{P}_j \right)$ is the set of the state variables separately either $j=\theta$ or $\ell$. $\theta$ represents the angle and $\ell$ represents the bond length. $\bf{Q}_j$ and $\bf{P}_j$ are the coordinate and momentum variable sets for the system defined along $j$, respectively. When $h_{s}$ is the Hamiltonian for the target system, then the rest of the dynamics of the total system includes the thermal environment $\mathcal{E}$, which is completely independent of $x_j$. $h_{SE}$ is the interaction term between the target system and the environment. Therefore, $H_{\mathcal{E}}$ defined with its canonical variable set, $y_j =\left( \bf{Q}-\bf{Q}_{j},\bf{P}- \bf{P}_{j} \right)$ when $\bf{Q}$ and $\bf{P}$ indicate the total set of state variables. $PV_{\mathcal{E}}$ in Eq.(\ref{eqn:ensemble}) is the amount of energy given by the control factor from the outside of the system.
   
The interaction energy between the target system and the environment in $h_{SE}$ can be regarded as a strongly coupled system because it is derived from the limited number of particles in the adiabatic condition. When the target system has its variable defined as $x_\theta$, the environment's variable is $x_\ell$. The interaction between $x_\theta$ and $x_\ell$ as described using the correlated Cartesian coordinate system in Eq.(\ref{eq:eq4}) and Eq.(\ref{eq:eq5}) forms $h_{SE}$ as a strongly coupled bi-Hamiltonian.  The Hamiltonian that is affected by both $x_{\ell}$ and $x_{\theta}$ constructs the bi-Hamiltonian structure as noted by $h_{SE}$ in Eq.(\ref{eqn:ensemble}). 
 
 Following the stochastic thermodynamics frameworks suggested by Jarzynski\cite{Jarzynski2017}, $\phi(x_j)$ is the perturbation from the interconnection with the thermal environment, which is written as $\Delta H$ in Eq.(\ref{eq:eq_perturb1})$\sim$Eq.(\ref{eq:eq_perturb2}). The details of the derivation are included in Appendix B for the integrity of the derivation.

\subsection{Evolution of Cross Correlation under Smoluchowski picture }

In the atomic scale system, we presume that the small scale perturbation of the atoms, which is averaged close to zero, resolves the cross correlated condition. Since such a condition is regarded as not producing additional momentum in a group of particles, the evolution equation for cross-correlated states can be under the overdamping condition and is given by the Smoluchowski equation. Then, the evolution of the probability density function has its governing equation as follows:

\begin{align}\label{eq:eq_op}
\partial_t \rho = L_{C} \rho + L_{D} \rho + L_{S} \rho. 
\end{align}

$L_{C}$ is the Liouville operator with conservative force from potential energy. $L_{D}$ is the Fokker-Plank operator for the dissipative and random force. Lastly, $L_{S}$ is the Smoluchowski operator. The evolution of state variables governed by the Smoluchowski equation becomes 

\begin{align}
\label{eqn:smolu}
\xi \frac{\partial \rho}{\partial t} = \frac{\partial }{\partial x} \left(\frac{\partial \mathcal{U}}{\partial x} \rho  \right) + \frac{1}{\beta} \frac{\partial ^2 \rho}{\partial x^2}.
\end{align}

Here, $\xi$ is the damping coefficient. $\mathcal{U}$ is the potential function. 

It is supposed that the rearrangement of atoms in the target system does not induce any motion of the group of atoms. Therefore, the result of the master equation should satisfy $\xi \frac{\partial \rho}{\partial t} =0$ in Eq.(\ref{eqn:smolu}). Further details of the derivation for the substitution of Eq. (\ref{eqn:ensemble}) to Eq.(\ref{eqn:smolu}) is  in Appendix C. From the modified Hamiltonian with the perturbation $\phi$, the equation of motion(EOM) becomes

\begin{align}
\dot{\mathbf{Q}}_{j} = \frac{\partial H_{j}}{\partial \mathbf{P}_{j} }, \label{eq:eq_go1} \\
\dot{\mathbf{P}}_{j} =- \frac{\partial H_{j}}{\partial \mathbf{Q}_{j}},\label{eq:eq_go2} \\
\dot{\phi}= \frac{1}{\xi}\frac{\partial^2 \phi}{\partial x^2}, \label{eq:eq_cro}
\end{align}

with $j$ which is either $\ell$ or $\theta$. $H_j$ is as defined in Eq.(\ref{eq:eq_perturb1}) and Eq.(\ref{eq:eq_perturb2}) with $\Delta H$ that is equilvalent to $\phi$, which is not a mere perturbation term, but a term that limit the eigenstate for conservation of momentum. Note that the quantificaiton of $\phi$ based on the Dzhanibekov effect and its usage in diffusion process is well introduced in Appendix C and Appendix D, as well. 

The CGMD simulation using Eq.(\ref{eq:eq_go1})$\sim$ Eq.(\ref{eq:eq_cro}) proved its capability to replicate the nonlinear motion of the semi-flexible polymer system in previous studies\cite{Koh2021, Koh2023}. In this paper, the configuration in far-from-equilibrium conditions is adapted for further validation.

\section{Results}

\subsection{First Born approximation from EOM}
When the equation of motion is decided with the perturbation that is governed by Laplacian as shown in Eq.(\ref{eq:eq_go1})$\sim$Eq.(\ref{eq:eq_cro}), the eigenfunciton of the EOM can satisfy the conditions written as below:    
\begin{align}
\left( \frac{\hbar^2}{2m} \nabla^2+ \mathcal{U}(\vec{r}) \right) \Psi_0(\vec{r})=\frac{\hbar^2}{2m}  k^2 \Psi_0(\vec{r}), \label{eq:eq_laplacian1}\\
 \left(-\frac{\hbar^2}{2m} \nabla^2+ \mathcal{U}(\vec{r}) \right) \Psi_1(\vec{r})=\frac{\hbar^2}{2m}  k^2 \Psi_1(\vec{r}),\label{eq:eq_laplacian2}
\end{align}

with 
\begin{align}
 \sum_j \nabla^2 \mathcal{A}_{0,j}\Psi_{0,k} = - \sum_j \nabla^2 \mathcal{A}_{1,j}\Psi_{1,j}.
 \end{align}

  $\Psi_{i}$ with $i=0$ or $1$ is eigenfunction. $\Delta t$ is from the integration of Eq.(\ref{eq:eq_cro}), and $\mathcal{U}(\vec{r})$ is potential energy function which is defined as harmonic. $\vec{r}$ is the vector drawn from the origin to the mass point in the tube in spherical coordinates. $\mathcal{A}_{i,j}$ is the amplitude for the eigenfunction $\Psi_i$ with $j$ th eigenvalue. $\hbar$ is planck constant which should be considered as effective constant from normalizaiton with $\frac{\Delta t}{\xi}$, and $\xi$ is a parameter for Eq.(\ref{eq:eq_cro}).  Since the target system is defined either by the Hamiltonian for bond length, $\ell$, or the Hamiltonian for angle, $\theta$, that share the perturbation with opposite signs, the integral equation for the time-independent form is valid when two pairs of EOM are considered together.  The sign of the perturbation for each type of Hamiltonian is flexible as written in Eq.(\ref{eq:eq_perturb1}) and Eq.(\ref{eq:eq_perturb2}), and it is supposed to be affected by the spontaniously given entropy production rate from the perturbation. 
  
  The eigenfunction of each harmonic Hamiltonian is as what have been considered with diagonalization. The Laplacian in Eq.(\ref{eq:eq_cro}) defined along the EOM, however, transforms the eigenstate from harmonic Hamiltonian to be the solution in Eq.(\ref{eq:eq_laplacian1})$\sim$Eq.(\ref{eq:eq_laplacian2}). $\Psi_i$ with $i=0$ or $1$ is not identical to the eigenfunction of harmonic Hamiltonian especially when the solution of Eq.(\ref{eq:eq_laplacian1})$\sim$Eq.(\ref{eq:eq_laplacian2}), is estimated with first Born approximation in following paragraphs. However, it is mentioned as so because it is considered with the wavevector that the simple beads system connected in series could have as introduced in any related textbook for our convience of quantification in following subsections. For the goal of distinguishing the scattering from the Laplacian, it is enough to quantify as the modification of the eigenfunction with given wavevector. Any set of eigenfunction of the EOM in Eq.(\ref{eq:eq_go1})$\sim$Eq.(\ref{eq:eq_cro}) would satisfy or closely approxiamted through the scattering function that we are going to introduce.    

Then, the eigenfunctions for each governing equation in Eq.(\ref{eq:eq_laplacian1}) and Eq.(\ref{eq:eq_laplacian2}) becomes 

\begin{align}
\Psi_{0,\pm}\left( \vec{r} \right)= e^{\pm i k x} + \int d^3 \vec{r}' G_{0,\pm}(\vec{r}-\vec{r}') \mathcal{U}(\vec{r}') \Psi_{0,\pm}( \vec{r}'), \label{eq:eq_eigenfunction1}\\
\Psi_{1,\pm} \left( \vec{r} \right)= e^{\pm  k x} + 
\int d^3 \vec{r'} G_{1,\pm}(\vec{r}-\vec{r}') \mathcal{U}(\vec{r}') \Psi_{1,\pm}( \vec{r}'). \label{eq:eq_eigenfunction2}
\end{align}

 The first term for $\Psi_{0,\pm}$ and $\Psi_{1,\pm}$ is homogeneous analystical solution that is defined along the tube axis $x$, and the second term with integration is defined with the Green function, $G_{\pm,i}$ with $i=0$ or $i=1$ for each definition in Eq.(\ref{eq:eq_eigenfunction1}) and Eq.(\ref{eq:eq_eigenfunction2}). The eigenfunction defined in Eq. (\ref{eq:eq_eigenfunction2}) is supposed to compensate for the scattered energy in Eq. (\ref{eq:eq_eigenfunction1}) in the integration.
 
  The alternation of the signs corresponds to a scattering state that can be examined under the far-from-equilibrium conditions prepared in the simulation for the collision between SWCNTs, which is conducted using an atomic-scale molecular dynamics (MD) simulation. As shown in Fig. 2A, two (5,5) SWCNTs with a length of 20 nm are arranged at 90 degrees to intersect at the midpoint of the tube. To induce the collision, Tube 2 in Fig. 2 is artificially bent with a 5 nm depth and released to make a collision with Tube 1. The collision response is as shown in Fig. 2B. From the clear decay of the bending mode from the collision, the amplitude of the eigenfunction with the first bending mode in Eq.(\ref{eq:eq_eigenfunction1})  is measured from normal mode decomposition using the trajectory of the simulation after it is averaged with 50 atoms and is trasformed into a numerically driven coarse-grained(CG) model for normal mode decomposition. From the decay in Fig. 2B, the eigenstate with the negative sign in Eq.(\ref{eq:eq_eigenfunction2}) are expected to dominate.
 

With the first Born approximation, the integration in Eq.(\ref{eq:eq_eigenfunction1}) and Eq.(\ref{eq:eq_eigenfunction2}) can be written with the scattering funciton $f_{0,\pm} $ and $f_{1,\pm} $ as below:

\begin{eqnarray}
f_{0,\pm} =\int d^3 \vec{r}' G_{0,\pm}(\vec{r}-\vec{r}') \mathcal{U}(\vec{r}') \Psi_{0,\pm}( \vec{r}') \nonumber \\
  = \int d^3\vec{r}' \frac{e^{\pm ik |\vec{r}-\vec{r}'|}}{|\vec{r}-\vec{r}'|} \mathcal{U}(\vec{r}')e^{\pm i \vec{k} \vec{r}'}, \label{eq:eq_scattering1} \\ 
f_{1,\pm} = \int d^3 \vec{r'} G_{1,\pm}(\vec{r}-\vec{r}') \mathcal{U}(\vec{r}') \Psi_{1,\pm}( \vec{r}'), \nonumber \\ 
=\int d^3\vec{r}' \frac{e^{\pm k |\vec{r}-\vec{r}'|}}{|\vec{r}-\vec{r}'|} \mathcal{U}(\vec{r}')e^{\pm \vec{k} \vec{r}'}. \label{eq:eq_scattering2}  
\end{eqnarray}

  \begin{figure}
 \includegraphics[scale=0.5,trim={0cm 11cm 20cm 0cm},clip]{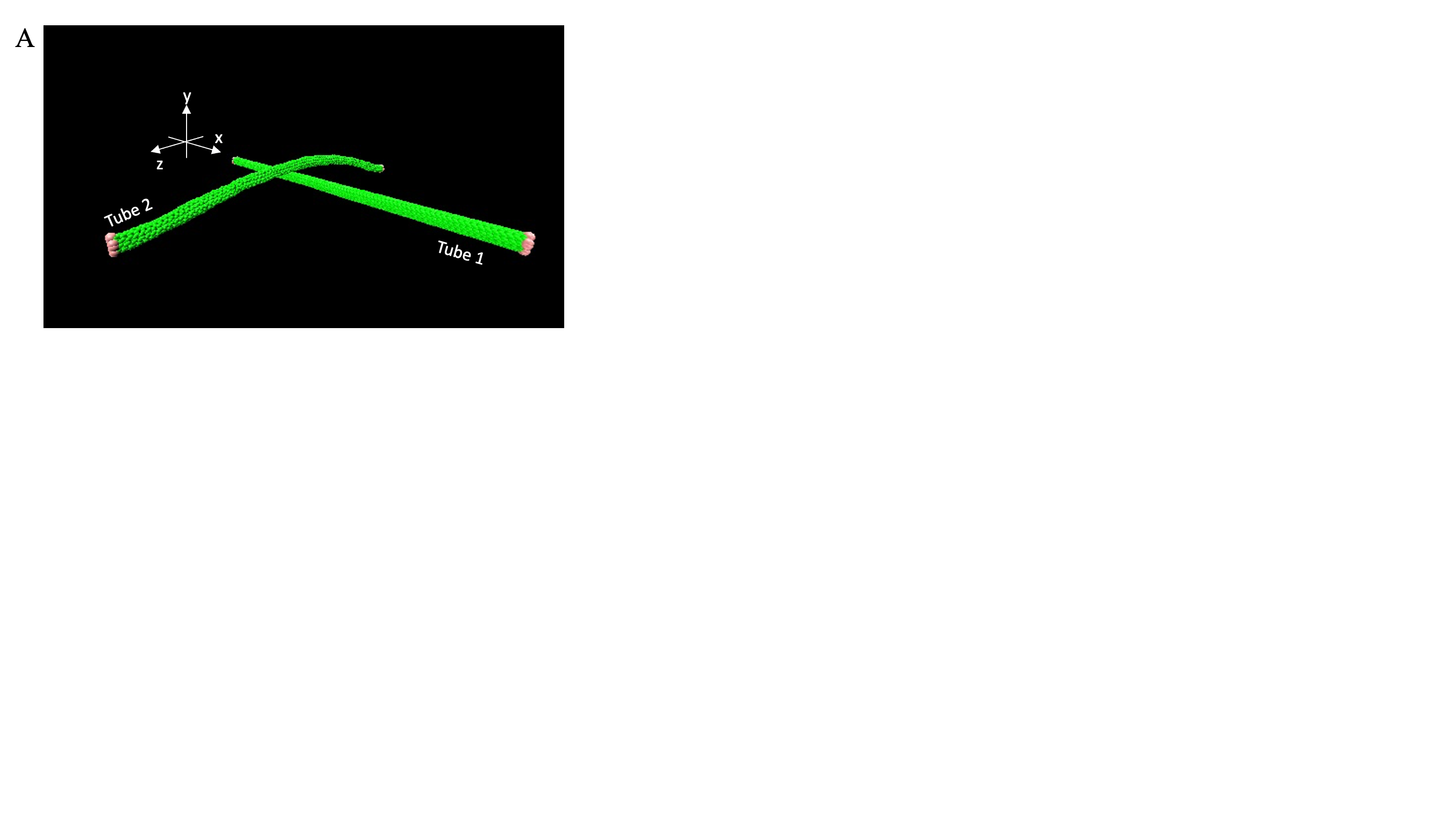}
  \includegraphics[scale=0.35,trim={0.0cm 0cm 1cm 1cm},clip]{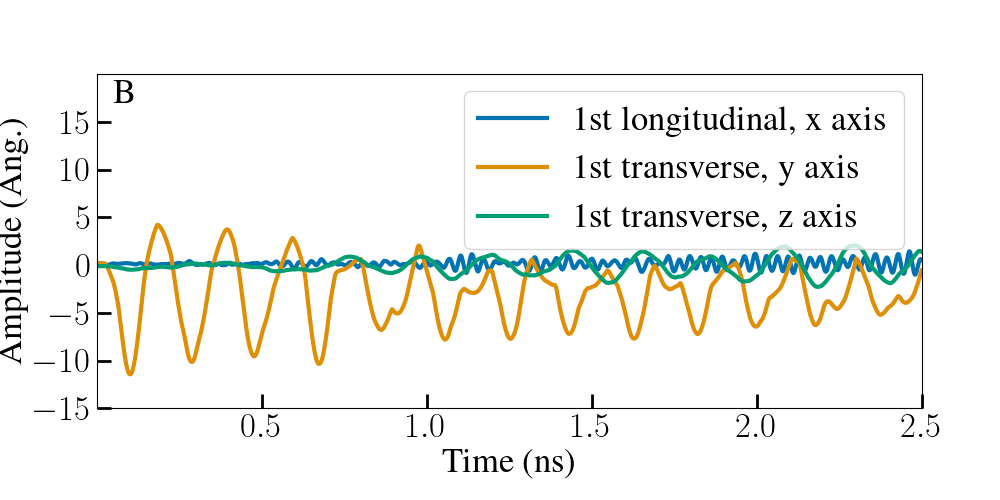}
 \caption{A. The initial configuration of the molecular dynamics(MD) simulation for the collision between two (5,5) SWCNTs. Tube2 is artificially bent before the simulation and its trajectory is disregarded in the analysis. B. The data from Tube1, which receives the collision, are collected for normal mode decomposition and subsequent analysis. The first mode along each Cartesian coordinate is calculated for the trajectory from the MD simulation.}
 \end{figure}

  \begin{figure}
    \includegraphics[scale=0.35,trim={0.5cm 0cm 0cm 0cm},clip]{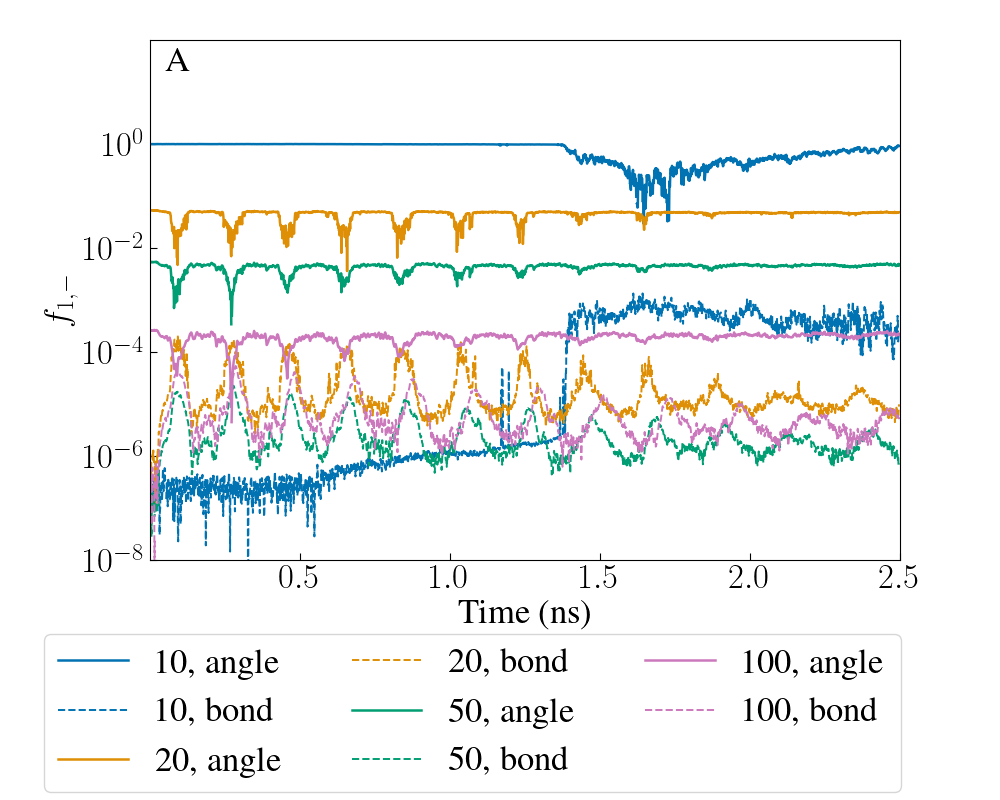}
    \includegraphics[scale=0.35,trim={0.5cm 0cm 0cm 0cm},clip]{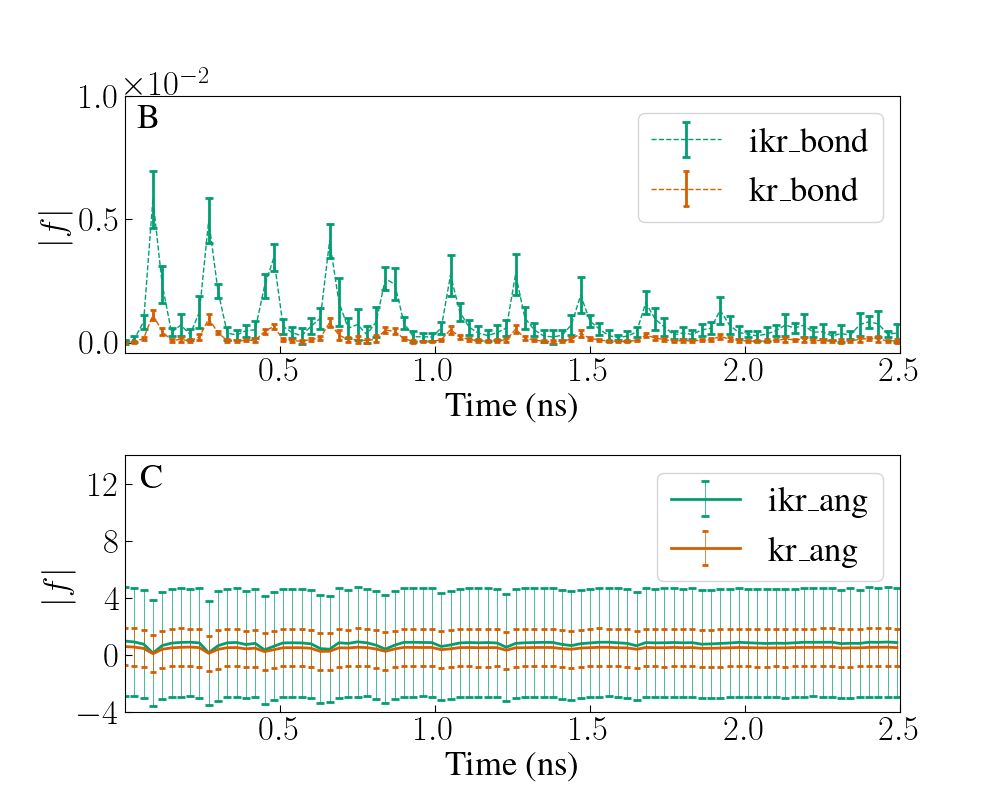}
 \caption{A. Scattering function from eigenstate $\Psi_1$ for different coarse-grained particles, averaged from the Tube 1 in MD simulation trajectory, B. Amplitude of scaterring function derived from the eigenfunction $\Psi_0$ and $\Psi_1$ for 1st mode of bond length potential and angle potential cases of the numerically driven CG model averaged 50 atoms for a particle. }
 \end{figure} 
 
Note that the integration is defined along the tube so that the vector $\vec{r}-\vec{r}'$ in spherical coordinates becomes the measure along the tube, and the wave vector or decay rate along the distance, $k$ is given with $x$ which is in parallel with the vector $\vec{r}-\vec{r}'$ in Eq.(\ref{eq:eq_eigenfunction1}) and Eq.(\ref{eq:eq_eigenfunction2}). 

Quantification of Eq. (\ref{eq:eq_scattering1}) and Eq. (\ref{eq:eq_scattering2}) is conducted through the trajectory data of MD simulation to validate the usage of the Laplacian in the EOM in Eq.(\ref{eq:eq_go1})$\sim$Eq.(\ref{eq:eq_cro}) which also justifies the implementatino of Laplacian for the EOM of CG system. The trajectory data from the result of the MD simulation shown in Fig.2 are translated into a numerically driven CG model by averaging the coordinates of a group of atoms to offer the potential energy landscape. For different particle sizes, $f_{1,-}$ is calculated using bond length and angle potential energy, $\mathcal{U}_b$ and $\mathcal{U}_a$, respectively. The parameter $k$ in $e^{\pm kx}$ in Eq.(\ref{eq:eq_scattering2}) is regarded as the decay distance and is identically designated with the wave vector $k$ in $e^{\pm ikx}$, which is from the possible wave vector of the numerically driven CG model. 

In Fig. 3A, the mean value of each scattering function along the tube for $f_{1,-}$ in Eq. (\ref{eq:eq_scattering2}) during 2.5 ns is shown. When 10 atoms are averaged into coarse grained particle, there is no meaningful scattering process that is corresponding to the collision process as shwon in Fig. 2B. Larger CG particles, which are averated with 20, 50 and 100 atoms, the scattering function in Eq. (\ref{eq:eq_scattering2}) has the response that is agreable with the trajectory analyzed with normal mode decomposition for 1st mode, $k_1$. The scattering functions calculated with $\mathcal{U}_b$ are at a relatively lower level than those calculated with $\mathcal{U}_a$. For every case except the averaging 10 atoms per CG particle, for each type of potential energy, have peaks or notches that correspond to the collision between the tubes.  

In the case of $f_{0,+}$ from Eq. (\ref{eq:eq_scattering1}), a similar response is observed with the trajectory data that is averaged with 50 atoms per CG particle, as shown in Fig. 3B and Fig. 3C with the error bar for deviation. The scattering function for $\Psi_0$ and $\Psi_1$ for $\mathcal{U}_b$ in Fig. 3B has clear peaks compared to the case quantified with $\mathcal{U}_a$, whose deviation is rather large compared to the mean value of the scattering function as shown in Fig. 3C. $f_{0,-}$ is not evaluated separately becuase it has the off phase condition from $f_{0,+}$. This difference can be revealed through the phase of $f_{0,+}$, which is introduced in the next subsection.

\subsection{Phase distribution of scattering function}

  \begin{figure}
  \includegraphics[scale=0.35,trim={0.0cm 0cm 0.0cm 0cm},clip]{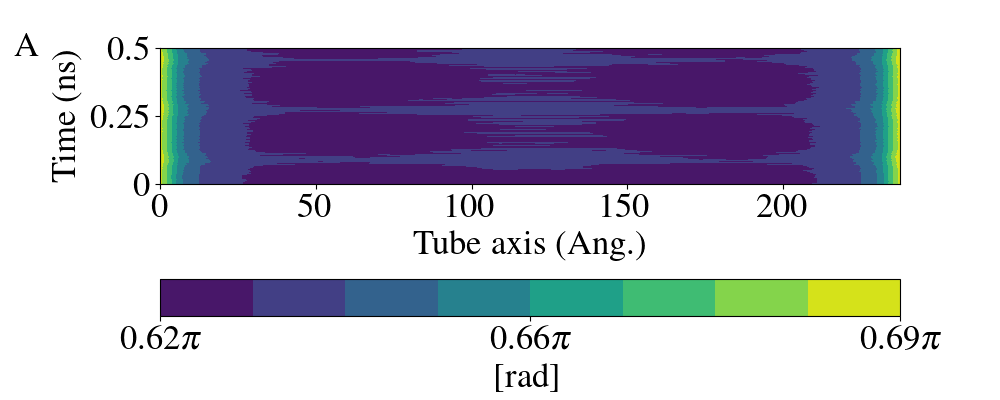}
   \includegraphics[scale=0.35,trim={0.0cm 0cm 0.0cm 0cm},clip]{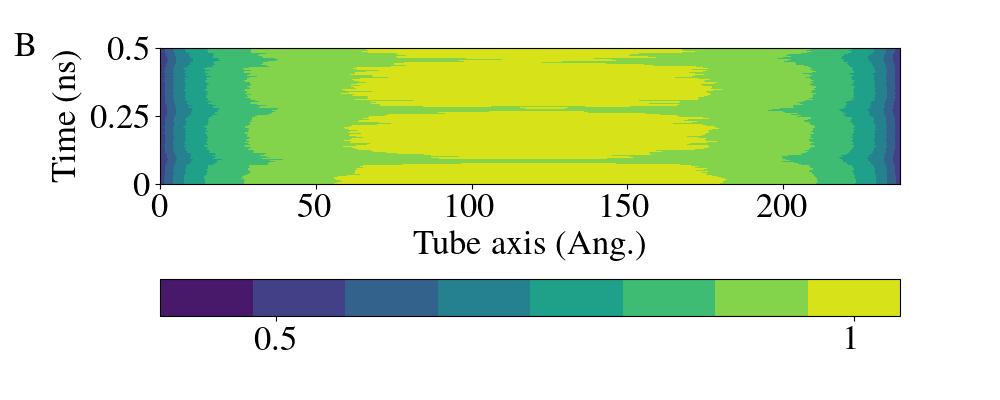}
   \includegraphics[scale=0.35,trim={0.0cm 0cm 0.0cm 0cm},clip]{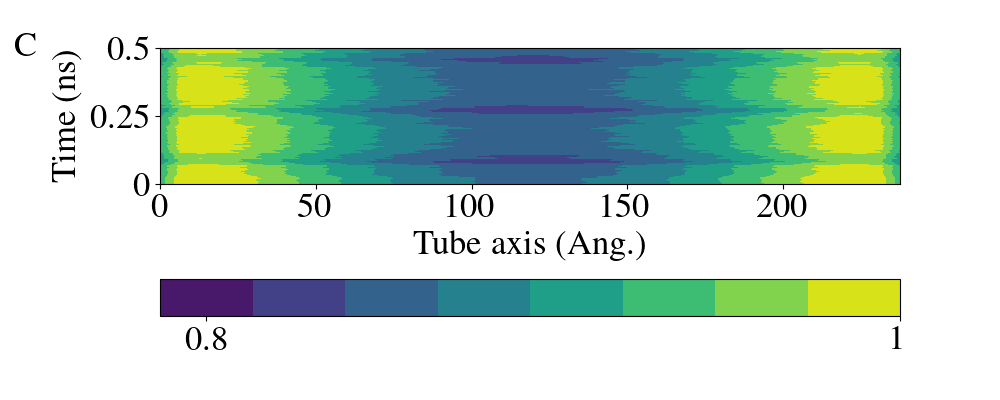}
 \caption{A. Phase distribution measured through the scattering function, $f_{0,+}$ from the potential energy of the angle variable calculated from the trajectory of the MD simulation given in Fig. 2 for the first bending mode, B. Distribution of the scattering function, $f_{1,-}$, with the relaxation distance defined from the first bending mode wavelength, C. Distribution of $f_{1,+}$ calculated with $\mathcal{U}_{a}$ and the relaxation distance defined from the first bending mode wavelength. }
 \end{figure}

   \begin{figure}
  \includegraphics[scale=0.35,trim={0cm 0cm 0cm 0cm},clip]{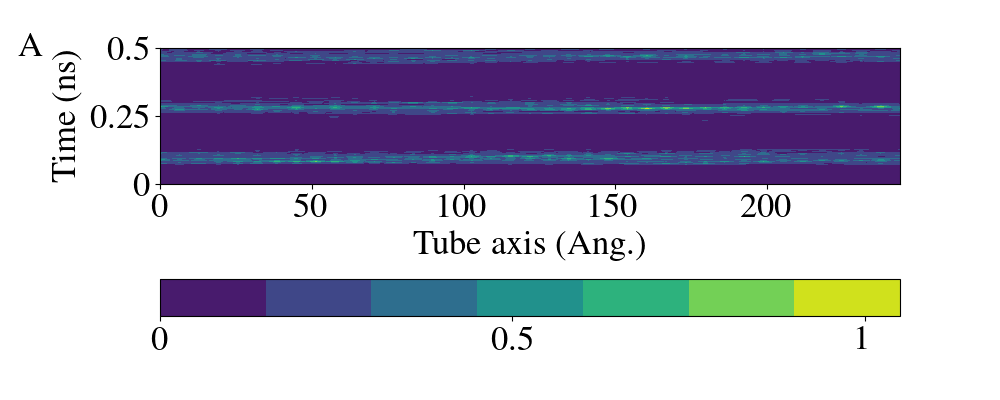}
    \includegraphics[scale=0.35,trim={0cm 0cm 0cm 0cm},clip]{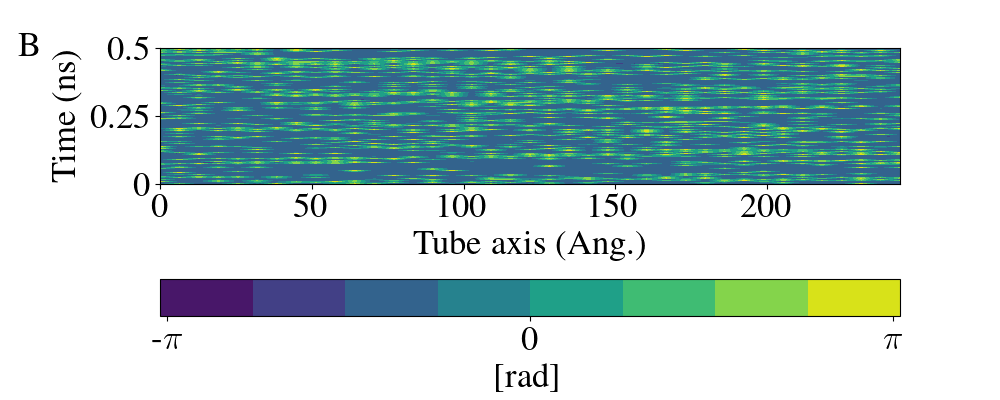}
      \includegraphics[scale=0.35,trim={0cm 0cm 0cm 0cm},clip]{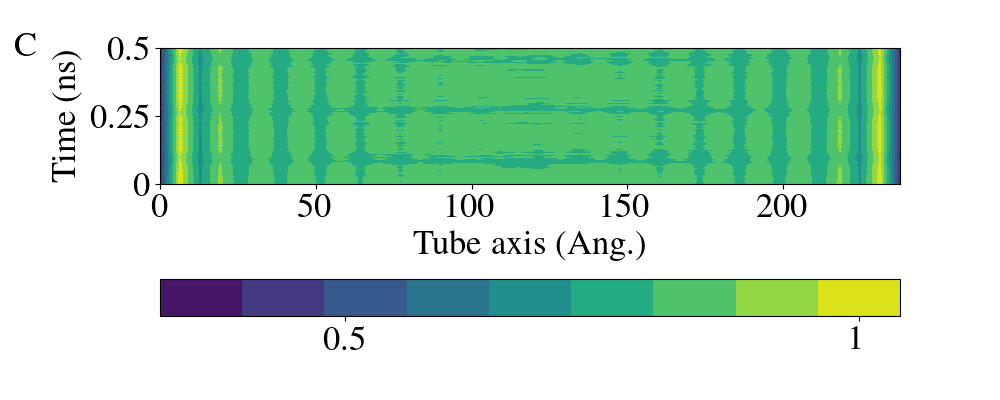}
    \includegraphics[scale=0.35,trim={0cm 0cm 0cm 0cm},clip]{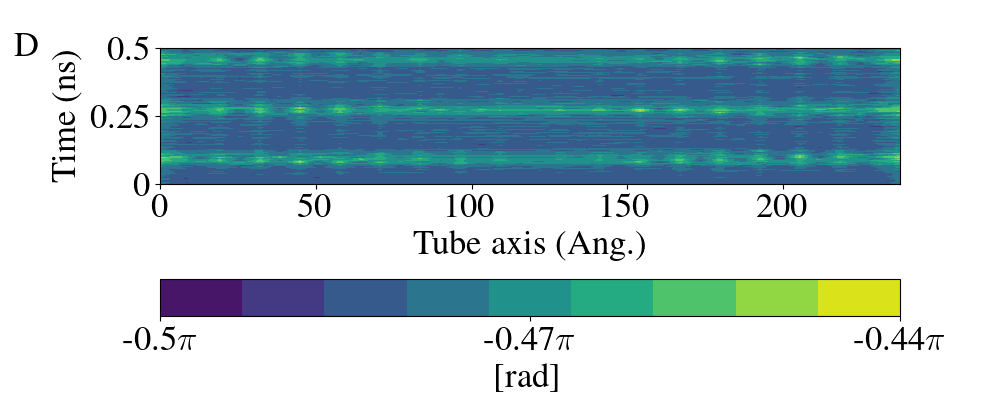}
 \caption{ A.and B. Amplitude and phase distribution of $f_{0}$ the Nth longitudinal mode from the bond-length potential energy, C and D. Amplitude and phase distribution of $f_{0}$ from the potential energy of the angle variable calculated from the trajectory of the MD simulation given in Fig. 2 for the Nth bending mode.}
 \end{figure}

The phase distribution of $f_{0,+}$ for $\mathcal{U}_a$ with first bending mode wave vector, $k_1$ is in Fig. 4A. Altogether with its large deviation on the averged amplitude along the tube as shown in Fig. 3C, the result of phase distribution corresponding to the collition time indicate the deformation of the tube is scattered. The distribution of $f_{1,\pm}$ of $\mathcal{U}_a$ with $k_1$ normalized to the maximum value follows what is observed in $f_{0,+}$ with an identical pattern. Yet, their diminished amplitudes indicate that the decay given by $e^{-kx}$ or the generation of the mode which is projected with $e^{kx}$ is slightly suppressed. Therefore, it is not the eigenstate $\Psi_1$ projected with $\mathcal{U}_a$ and $k_1$ that is responsible for the decay of the collision response. It is the scattering function $f_{1,\pm}$ of $\mathcal{U}_b$ that induces the proper decay of collision energy and the generation of local deformation corresponding to the collision at the same time.  
 
 Meanwhile, there is the result of the $N$ th mode wave vector, $k_n$ that proves the eigenstates $\Psi_0$ corresponding to the pattern of neceesary decay of the collision response as shown in Fig. 5. What to notice is that the dynamics with $k_n$ shows a distinctive pattern of fluctuation demonstrated even before the collision during 0 ns $\sim$ 1 ns in Fig. 5B and C. The phase from $\mathcal{U}_b$ in Fig. 5B and the amplitude from $\mathcal{U}_a$ in Fig. 5C imprint those intrinsic pattern during the collision process. More detail analysis on the distinctive pattern that has shown in Fig. 5A and Fig. 5D is in Discussion section.  
 
 Finally, the scattering functions, $f_{1,\pm}$ for the eigenfunction $\Psi_1$ with $k_n$ are observed in Fig. 6.  $f_{0,-}$ for $\mathcal{U}_a$ using N the mode decay distance, $k_n$, the similar surpression of the amplitude during the collision is observed as shown in Fig. 4B and 4C. It is the scattering function $f_{1,-}$ for $\mathcal{U}_b$ calculated from $k_n$ that corresponds to the decay of the collision energy in Fig. 6B. Clear peaks distributed along the tube length is observed in Fig. 6B with a short dashed patterns as observed in Fig. 5B, following the collision occurrence of the tube as shown in Fig. 2B. The range of $f_{1,-}$ with corresponding to the collision process as shown in Fig. 6B indicates that the decay expressed by the eigenstate $\Psi_1$ with $e^{-kx}$ is responding to what we observed in Fig. 5A, which is calculated with $e^{-ik_{n}x}$. Therefore, the analyse through the data processed with scattering function indicates the small wavelength fluctuation that makes the given dynamics of the tube responds to the decay from the collision, and the correlated response is lead to a short decay distance attenuation described with $\Psi_1$.

   \begin{figure}
  \includegraphics[scale=0.35,trim={0cm 0cm 0cm 0cm},clip]{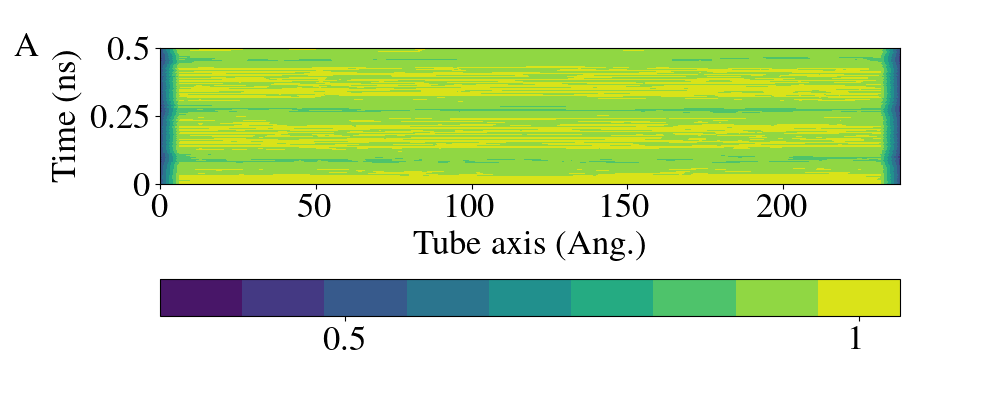}
  \includegraphics[scale=0.35,trim={0cm 0cm 0cm 0cm},clip]{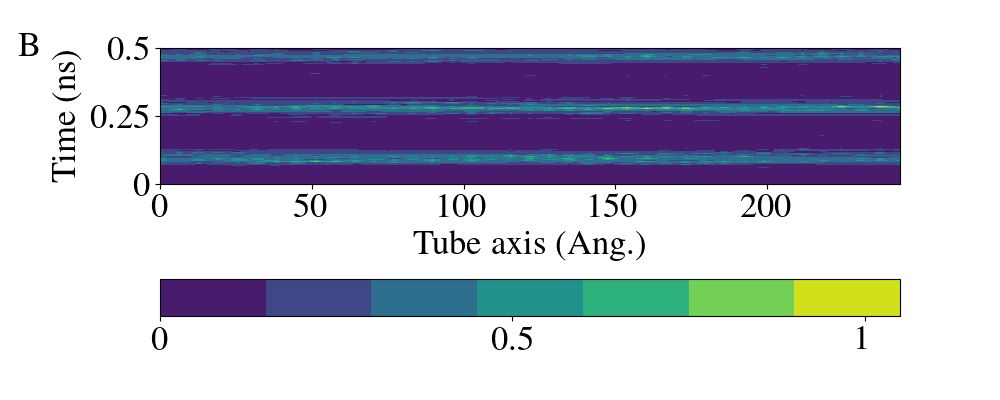}
 \caption{ Distribution of scattering function, $f_{1,-}$ with $k_n$. A. calculated with $\mathcal{U}_a$ with the relaxation distance defined from N th mode wavelength, B. calculated with $\mathcal{U}_b$ with the relaxation distance defined from N th mode wavelength.}
 \end{figure}

\subsection{Coarse grained simulation}
 
  \begin{figure}
 \includegraphics[scale=0.5,trim={0cm 13cm 20cm 0cm},clip]{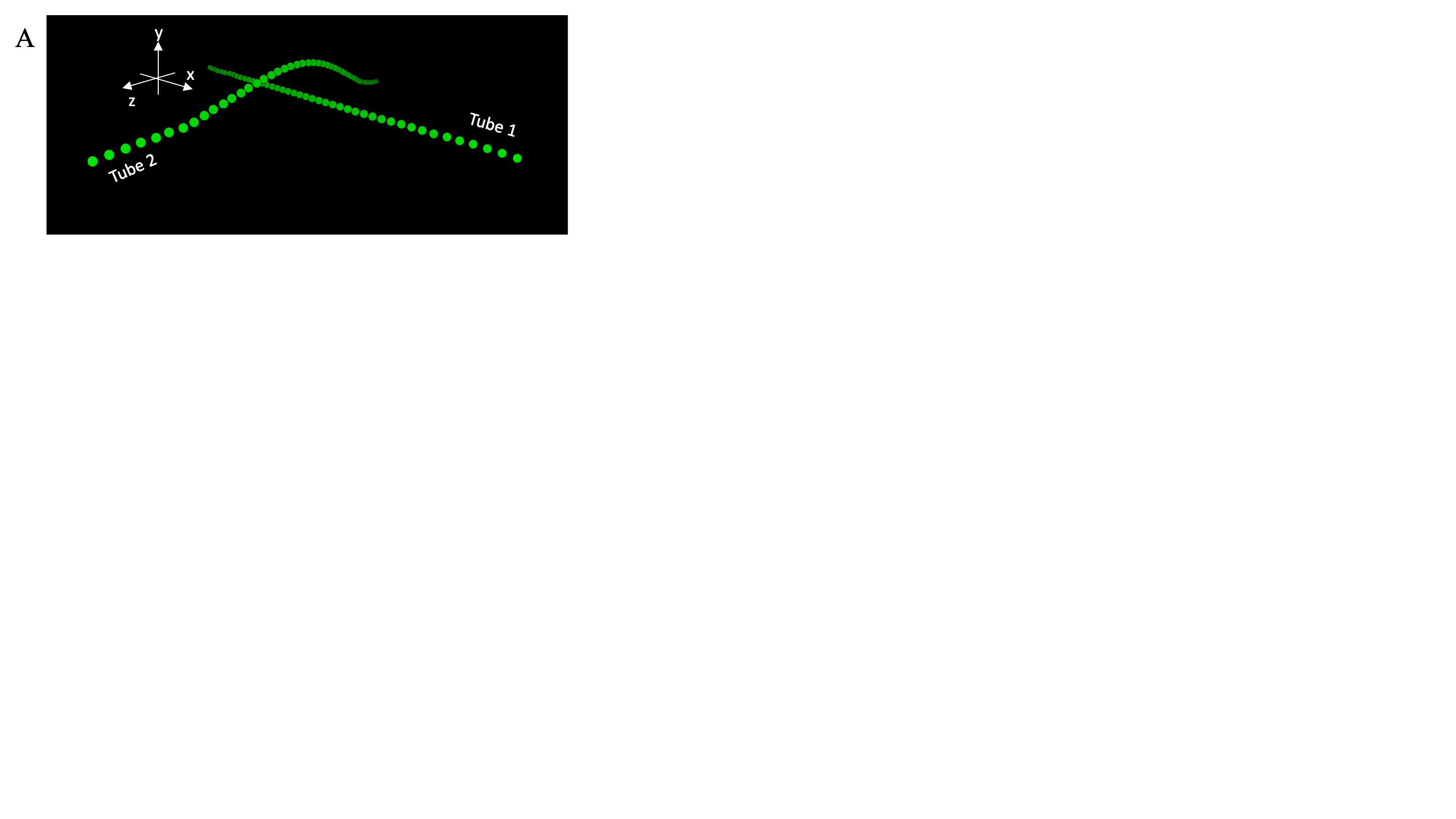}
  \includegraphics[scale=0.35,trim={1cm 0cm 0cm 0cm},clip]{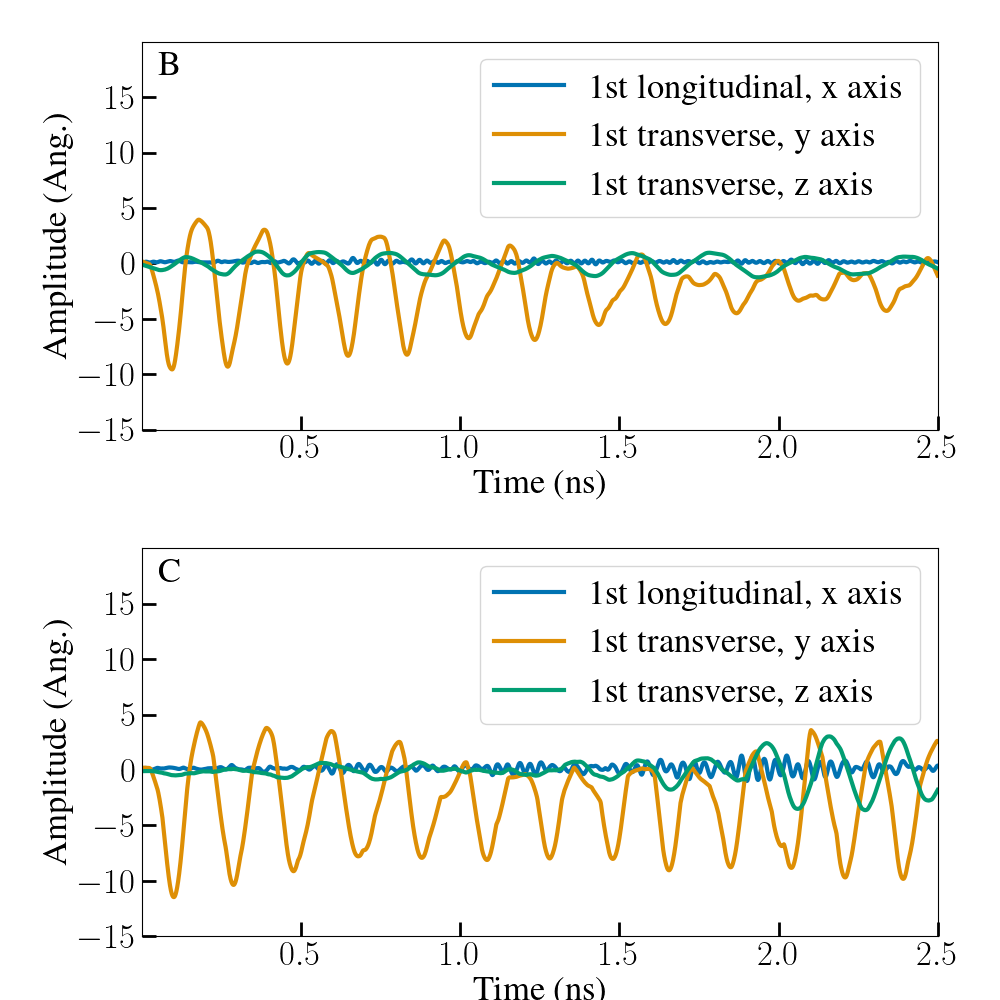}
 \caption{A. The initial configuration of CGMD model. Tube2 is artificially bent before the simulation, and the initical condition is calculated from MD simulation model in Fig. 2A. The data from Tube1, which receives the collision, are collected for normal mode decomposition and subsequent analysis.  The result of normal mode decomposition. B. CGMD simulation with Eq.(\ref{eq:eq_go1})$\sim$Eq.(\ref{eq:eq_cro}), and C. CGMD simulation without Eq.(\ref{eq:eq_cro}). }
 \end{figure}
   \begin{figure}
  \includegraphics[scale=0.35,trim={1cm 0cm 0cm 0cm},clip]{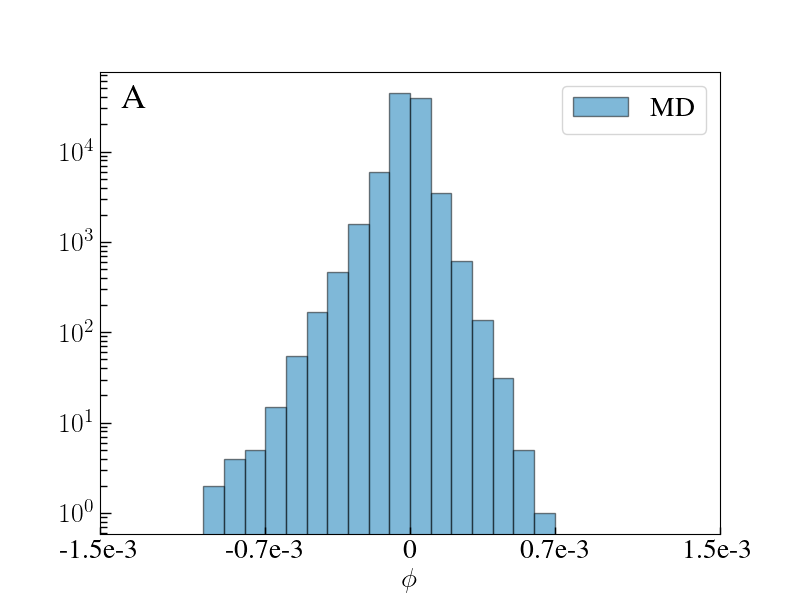}
  \includegraphics[scale=0.35,trim={1cm 0cm 0cm 0cm},clip]{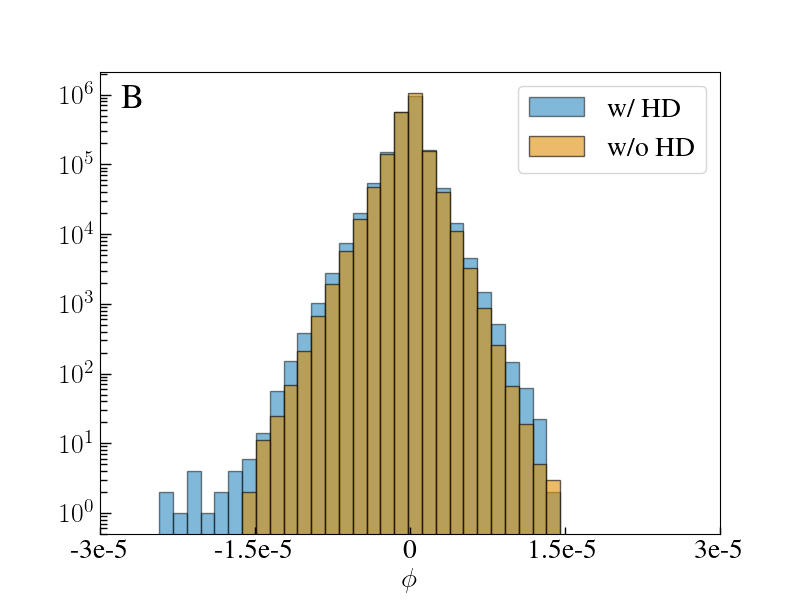}
 \caption{The histogram of $\phi$  A. in the MD simulation, B. in the CGMD simulation with(marked as w/ HD in blue shade) and without heat diffusion process(marked as w/o HD with yellow shade) that is derived as Eq.(\ref{eq:eq_cro}).}
 \end{figure}
 
 From the decay process observed using scattering function $f_{1,-}$, which is derived with $k_n$ for $\mathcal{U}_b$ and the result from $f_{0,+}$ in Fig. 5, it is clear that the short wave length is activated for decay of the collision energy for the relaxation shown in Fig. 2B. From this result in previous subsection, the diffusion process derived from Smoluchowski equation in Eq.(\ref{eq:eq_cro}) in CGMD simulation is implemented with short wave length fluctuation mimicking the inherited dynamcis of the tube as shown in Fig. 5A and C. It is straightforward because the calculation of the diffusion equation and the fluctuation from short wavelengths share the same numerical derivation in a discretized system. Therefore, adjusting the CG system to short-wavelength fluctuations, as shown in Fig. 5, is expected to be equivalent to applying the diffusion process observed in Fig. 6.  
 
 The data from the MD simulation is then converted into CG particle information by averaging the displacement and velocity of each group of 50 atoms per CG particle.  The CG model, as shown in Fig. 7A, is calculated using Eq.(\ref{eq:eq_go1})$\sim$Eq.(\ref{eq:eq_cro}). The quantification of $\phi$ during the simulation is in Appendix D, and the details of the simulations are in Appendix E. The 1st mode of deformation along each direction caused by the collision is as shown in Fig. 7B and 7C. The amplitude along y axis clearly proves that the CGMD simulation conducted with Eq.(\ref{eq:eq_go1})$\sim$Eq.(\ref{eq:eq_cro}) has better agreement with the result from MD simulation. The CGMD simulation with Eq. (\ref{eq:eq_cro}) has not recovered the amplitude of the flexural mode after 2.5 ns. This indicates that the mode coupling process, which attenuates the amplitude, has been successfully activated. Without the diffusion process in Eq. (\ref{eq:eq_cro}), as shown in Fig. 7C, relaxation of the collision energy does not occur. 

Another proof that the CGMD simulation with Eq.(\ref{eq:eq_cro}) has identical dynamics with the MD simulation is the histogram of the value of $\phi$ that is extracted from each CGMD and MD simulations as shown in Fig. 8. The $\phi$ collected from the numertically driven CG model from the MD simulation data has skewed symmetry as shown in Fig. 8A, which is from the the presence of attenuation. The same skewed shape is observed in the histogram from the CGMD simulation with the diffusion process, Eq.(\ref{eq:eq_cro}).  The $\phi$ exists even in the conventional CGMD simulation because the particle is under the influence of the Dzhanibekov effects. Yet, there is no damping caused by Eq.($\ref{eq:eq_cro}$); therefore, the shape of the histogram remained symmetrical. 

  Conceptually, the $\phi$ is regarded as the energy exchange between two Hamiltonians through fast dynamics. When fast dynamics preclude equal exchange across Hamiltonians and the negative value of $\phi$ is overpopulated, attenuation is inevitable. The skewed symmetry in Fig. 8A and 8B is the result of unequal exchange, leading to attenuation of the collision response, as shown in the MD simulation. The sampling frequency, however, could not be identical between MD and CGMD simulation which has artificially given fluctuation that mimic the N th mode dynamics of atomic scale model as shown in Fig. 5 and 6. This causes the population around $\phi \approx 0$ from CG simulation increased compared to that of MD simulation with more high sampling frequency as mentioned in the appendix E. Without such different sampling frequency, the skewed symmetry from CGMD simulation with Eq.(\ref{eq:eq_cro}) become faint.

\section{Discussion}

\subsection{Bi-Hamiltonian incarnation from overdamping process}
In the previous work on the nonlinear bending replcated by CGMD system and the collision process in this paper reproduced using identical simulation technique, at least two independent Hamiltonians is required to reconstruct the macroscpic motion of the atomic scale n body system for semiflexible polymer. Due to the adjunct process that relies on the overdamping and its result in diffusion equation, the dynamcis of suggested CG system seems not identically aligned with bi Hamiltonian structure. However, the trajectory from modified CGMD simulation with $\phi$ can be regarded as the reconstrucion of bi Hamiltonian or the result of bi Hamiltonian with two following perspectives. 

First, it is the two given indpendent CG Hamiltonians that offers the result of poisson tensor that decides the trajectory of the dynamics, judging from the conventional intuition on overdamping process that hardly produce the distinctive trajectory of its own. The missing aspect is that the result of trajectory modeled with two CG hamiltonian and an overdamping process does not offer another Hamiltonian structure of the CG system. What is reconstructed is the trajectory that is drawn by n-body Hamiltonian of the original system. We may not able to find a proper set of state variables that consitute the another Hamiltonian that covers two CG system with an overdamping process.  However, those harmonic Hamiltonians implemented in CG system is a part of the dynamics systems of the original. From the perspective of the original system, the macroscopic trajectory becomes the result of the bi-Hamiltonian structure balanced with fast dynamics or equivalently overdmaping process. 

Second, the reconstructed dynamics is under the scheme of integrability because of what the overdamping process offers for its eigenfunction, $\Psi_1$. With the potential energy defined along the coordinate variables expressed along the exponential term, the set of Hamiltonians that defines the dynamics of the integrable system is secured as a viable options. From the analogy of the fast dynamics which draws small perturbations to the main Hamiltonian that describes the macroscopic motion with additive poisson tensor in bi Hamiltonian pencil, the overdamping process and its possible integrability can be referred as a suitable compensation to the CG hamiltonians, as well as a solid perspective that the system of semiflexible polymer has a bi-Hamiltonian structure.

\subsection{Eigenstate for overdamping process and TUR}

The harmonic potential energy from the eigenstate $\Psi_1$ is in the form of $e^{-kx}$. This could induce an offset at $x=0$. However, the distribution is minimized when the displacement variables becomes large, so that it rarely disturbs the given harmonic potential energy; therefore, it is compatible with the given harmonic Hamiltonian and its variables, which are bond length, $\ell$, and angle $\theta$. Its alternation of plus and minus signs that is applied to $H_{\ell}$ and $H_{\theta}$ without duplication also bolsters the stability of the derived EOM with evanescent energy from the eigenfunction, $\Psi_1$. For $e^{\pm kx}$ in $\Psi_1$ manifesting the decay or evanescently generated deformation, it should be the diffusion process of $\phi$ in Eq.(\ref{eq:eq_cro}) that is described by $\Psi_1$. Furthermore, the eigenfunction $\Psi_1$ and the potential energy induced from it explains the shape of the histogram in log scale as shown in Fig. 8A and B. The difference is distinguished in the left side of the histogram, which indicates the range of $\phi$ with a minus sign has more population. The linear shape in log scale represents the eigenstate $\Psi_1$ with $e^{-kx}$, which is valid for the dynamics presented by $\phi$ in Eq.(\ref{eq:eq_cro}). 

Altogether with $\Psi_0$ and $\Psi_1$, we have the deviation and the median of the state variables that is described with the wave vector $k$, which is also regarded as the decay distance along the tube. Each information is the component that contistitue the definition of TUR. The information from the scattering functions, however, offers a slightly more enhanced understanding on the quantificaiton of TUR, because of the absence on the eigenvalue of each function. The entropy production rate, $\sigma_k$ from any state relying on the wave vector $k$ should be proportional to the following ratio: 

\begin{align}\label{eq:eq_entropy}
\sigma_k \ge \frac{2 \left<J\right>^2}{Var(J)} \propto \frac{ \left(1 + f_{1} \right)^2}{1+f_{0}}.
\end{align} 

 $J$ is any variable that represent the enegy flow from the wave vector $k$. Along the time span, it is the amplitude that we observed in NMD in Fig. 2B rather than the ratio between scattering functions from $\mathcal{U}_a$, $f_{0}$ and $f_{1}$, which assures the trend of $\sigma_{1}$. The surge of amplitude of the first mode from the collision, validates the subtancial change in variance. On the contrary,  $\sigma_{N}$ for $\mathcal{U}_b$ can be regarded as inhomogeous entropy production along the tube according to the shown in Fig. 5 and Fig. 6. This irregular entropy production along the tube bolsters the usage of diffusion process with the N th mode wave vector in $\phi$, the artificially adjusted perturbation in CGMD simulation, and its result from the simulation that identically follows the trajectory of SWCNT as observed in MD simulation. 

\subsection{Thermodynamics on Eq.(\ref{eq:eq_cro})}

 According to the definition of work and heat in Stochastic Thermodynamics, $\phi'$ as $\phi + w$ with the work $w$ done by the collision, is the heat energy that traverses the two distinguished Hamiltonians, affected by the external work $w$ . And the irregularity of scattering function in Fig. 5 and 6, the possibility of the heat energy exchange is discussed in previous subsection. Following the notation in VI Partial Molar Representation in Jarzynski's framework\cite{Jarzynski2017}, a certain amount of heat is induced when the $\phi'$ is the function of temperature as below:

\begin{align}
\phi_T = T \frac{\partial \phi'}{\partial T},\\
\bar{q} = q - \Delta \phi_T,\\
\bar{s} =  s-\frac{<\phi_T>^{eq}}{T}.
\end{align}

The temperature dependence of $\phi$ is clear from the Smoluchowski equation.  During the attenuation, it is regarded to be $\frac{\partial \phi}{\partial T} \neq 0$ and $\frac{\partial \phi}{\partial x} \neq 0$. The irregularity of $\phi$  along the tube can cause the unequal value of $\bar{s} $, so that this justifies the energy diffusion as the source of heat caused by the double partial derivative. 
 
 Determining how much entropy from $\phi$ contributes to the total entropy production and how each entropy production that is counted with a wave vector interacts remains difficult to be speicified at this early stage of research, without a rigorous study of generalization and parameterization. Inferring from another nonequilibrium steady state that is induced with a temperature gradient along the tube axis, the coarse-grained model seems not to include the heat transferred through the fast dynamics that is lost during the dimension reduction for coarse-graining, neither the influence of this fast dynamics with a certain amount of the heat flux to the macroscopic motion which is known as Z mode that enhances the heat conduction\cite{Koh2025}.  However, a more comprehensive exploration of the generalization of the definition $\phi$, particularly within specific mode-coupling or bi-Hamiltonian structures, is beneficial for excavating the quantities pertinent to TUR, especially when established within a stringent theoretical framework for its localized definition.

\section{Conclusion}

In previous work\cite{Koh2021}, the heat diffucion process from the memory effect caused by interaction between the target system and the bath is introduced as the possible replacement of the notorious integration over the memory effect presenting the fairly precise replication of the nonlinear bendimg motion of SWCNT in MD simulation using simple beads CGMD simulation. In this study, we show that it is the Dzhanibekov effect that causes the conflicts bewteen two independent Hamiltonian dynamics in CG model, and it is diffusion equation that resolves the conflict can be derived from overdamping process under Smoluchowski picture. The definition of the ensemble and its justification with relatively small number of particles is from the Stochastic Thermodynamics framework of Jarzynski\cite{Jarzynski2017}, and the derivation from Campisi\cite{Campisi2009a}. The derivation of diffusion process, substitutitng the ensemble into Smoluchowski equation follows the work presented back in 1997 by Espanol and Warren\cite{Espanol1995a}. 

As following, the integral equation from the EOM of two harmonic hamiltonian and the diffusion process validates the usage of the diffusion process in CG system, which has a slight twist with the condition in far from equilibrium with the collision process between tubes. The artificially implemented diffusion process is supposed to eliminate the cause of Dzhanibekov effect as balancing two independent Hamiltonian trajectory without breaking conservation law. To confirm the way of implementation of diffusion process in CGMD simulation and its rollout in the dynamics of semi flexible polymer, the scattering function of the integral equation is quantified through the first Born approximation and the numerically driven CG system from the trajectory information of MD simulation, which offers the potential energy landscape in identical resolution of CG system. From the distribution of scattering function, we could confirm the interplay of the modes with shorte wavelength as responding to decay or balance between two independent Hamiltonians that share the momentum of inertia. The implementation of diffusion process with the fluctuation over each time step and beads is also justified through the interpretation of two types of eigenfunctions in the definition of TUR. 

Starting with the analogy between the Hamiltonian of mean force and the memory effect that is derived from the bi Hamiltonian defined for the target system and the bath, the result of the derivation from the first Born approximation turns the perspective of the manuscript back into the bi Hamiltonian structure defined along the macroscopic motion of the semi fiexible polymer, the SWCNT whose macroscopic motion has almost identical replication through CGMD simulation composed with two independent Hamiltonian and an overdamping process. A more confirmative link to the TUR from the two species of eigenfunctions expands the scope of next study on characterizing the dynamcis in nonequilibrium.          



\bigskip

 \acknowledgements
This research is supported by Basic Science Research Program through the National Research Foundation of Korea(NRF) funded by the Ministry of Education (NRF-2022R1I1A1A01063582). Its computational resources are from National Supercomputing Center with supercomputing resources including technical support (KSC-2020-CRE-0345).  There are no conflicts to declare. The fortran code that is used to conduct the CGMD simulation of SWCNT collision is in \url{https://github.com/ieebon/SWCNT_collision}.

\appendix

\section{Finite Bath Fluctuation Theorem for SWCNT coarse graining}

Let the Hamiltonian of an atomic system of the heat bath be as below:

\begin{align}\label{eq:eq_Campisi1}
H_B(\{\vv{p}_{i} \},\{\vv{q}_{i} \}) = \sum^{n}_{i}\frac{\vv{p}_{i}^2}{2m} + \sum V(\{\vv{q}_{i} \}).
\end{align}

Following the derivation by Campisi et al.\cite{Campisi2009a}, the notations are retained as the same as the original reference. $\vv{p}_i$ and $\vv{q}_i$ are momentum and coordinate of the $i$th particle in the system. $\{\}$ indicates the set of the variables.

 The heat capacity of the bath is defined by the energy of the bath $E_B$ and the microcanonical temperature $T_B$ as follows:

\begin{align}
C(E_B) = \Biggl( \frac{\partial}{\partial E_{B}} T_{B}(E_{B})\Biggl)^{-1}, \\
T_{B} =\frac{ \Phi_{B}(E_{B})}{\Omega_{B}(E_B)} .
\end{align}

Here, $\Omega_B$ and $\Phi_{B}$ are  the density of states of the heat bath and the phase volume, respectively. For $C \rightarrow \infty $, there should be

\begin{align}\label{eq:eq_A4}
  \frac{\partial}{\partial E_{B}} T_{B}(E_{B}) = 1-\Phi_{B} \cdot \frac{1}{\Omega_{B}^{2}} \cdot \frac{\partial^2 \Phi_{B}(E_B)}{\partial E_{B}^{2}} \rightarrow 0,
\end{align}

when the bath density of the state and the phase space volume $\Phi_{B}$ satisfy

\begin{align}
 \frac{\Omega_{B}^{2}}{\Phi_{B}} \approx \frac{\partial^2 \Phi_{B}(E_B)}{\partial E_{B}^{2}}. 
\end{align}

According to Campsi et al\cite{Campisi2009a}, the phase space volume, $\Phi(E_B)_0$ for the ideal gas system in the box can be defined under the assumption that the phase volume of $\vv{q}_i$ is independent of $E_B$ as followings:

\begin{align}
\Phi_{B}(E_{B})_0 = \int \prod_{i=1}^{n}d\vv{q}_i \int \prod_{i=1}^{n} d\vv{p}_i \times \nonumber \\ 
\theta \Biggl( E_B - \sum^{n}_{i}\frac{\vv{p}_{i}^2}{2m} - \sum V(\{\vv{q}_i\}) \Biggl)   ,\\
= A_{dn}(2m)^{dn/2}E^{dn/2}_{B}V'^{n}
\end{align}

where $A_{dn} = \pi^{N/2}\Gamma(N/2+1) $ and $V'^{n} = \int \prod_{i=1}^{N} d\vv{q}_i$ for ideal gas of the hard sphere particles in volume $V'$ which is fixed. $\theta(x)$ is a heavyside step function. 

The system of interest in the main text is composed of hard spheres for CG particle that is connected as a single strand. The oscillation of the total system is conducted in a vacuum chamber, which has two interdependent Hamiltonians, $H_{\ell}$ and $H_{\theta}$. They have harmonic potential energy functions for bond length and angle, respectively. Therefore, the heat bath in relation to the Hamiltonian of the target system is redefined. When the target Hamiltonian has a harmonic potential energy for angle, the heat bath of that CG system has the Hamiltonian for bond length, and vice versa. 

In this context, a few variables in the Hamiltonian of the heat bath are defined as follows. The state variable in Eq. (\ref{eq:eq_Campisi1}) becomes the probablistic condition of $\mathbf{Q}_j$ and $\mathbf{P}_j$ for the coordinates and momenta, respectively. $j$ is either of $\theta$ or $\ell$. $\hat{P}$ indicates the momentum variable in the phase space. $E_B$ is the total energy of either of two Hamiltonians that the system has, which indicates that the level of $E_B$ is strongly coupled to the total energy of another Hamiltonian because the sum of those becomes a constant in the vacuum chamber. For this strongly coupled system defined as the harmonic oscialltors, the term $\int\prod_{i=1}^{N} d\hat{Q}_{i}$  is no longer independent to ${\hat{P}}$ or kinetic energy. For this issue, we can bring the interdependent condition between kinetic energy and potential energy that the sum of those quantities should be $E_B$ so that the value of $E_B$ in Eq. (B6) can be divided as $E_a$ and $E_B-E_a$ as below:

\begin{align}
\Phi_{B}(E_{B}) = \int \prod_{i=1}^{n}d\hat{Q}_{\alpha i} \int \prod_{i=1}^{n} d\hat{P}_{\alpha} \times \nonumber \\ 
\theta \Biggl( E_B - \sum^{n}_{i}\frac{\hat{P}_{i}}{2m} - \sum V_i(\{\hat{Q}\}) \Biggl),\label{eq:eq_A8}\\
= \prod dE_a \delta(\hat{E}_a-E_a) \times \nonumber \\
\int \prod_{i=1}^{n}d\hat{P}_{\alpha i}\theta \Biggl( (E_{B}-E_a) - \sum^{n}_{i}\frac{\hat{P}_{i}^2}{2m}  \Biggl)  \times \nonumber \\ 
\int \prod_{j=1}^{n} d\hat{Q}_{\alpha'_j} \theta \Biggl( E_a - \sum V_i(\{\hat{Q}\})  \Biggl),  \nonumber \\
\end{align}

here, $V(\{\hat{Q}_i\})$ indicates potential energy of the i th CG particle. Then the result of Eq. (\ref{eq:eq_A8}) becomes the following:

\begin{align}
\Phi_{B}(E_{B}) =  \prod dE_a\delta(\hat{E}_a-E_a) A_{dn}^2 (m/k)^{dn/2} \times \nonumber\\
(E_b-E_a)^{dn/2} E_a^{dn/2}\label{eq:eq_A10},
\end{align}
when $k$ is the spring constant for harmonic potential energy.  From this result of Eq. (\ref{eq:eq_A10}), the density of states of the bath $\Omega_B$,  $\frac{\Omega_B^2}{E_B}$ and $\frac{\partial^2 \Phi_B}{\partial E_B^2} $ which are the terms in Eq. (\ref{eq:eq_A4}) can be as follows: 

\begin{align}
\Omega_B(E_B) = \frac{\partial \Phi_B(E_B)}{\partial E_B} \nonumber\\
=dn/2 \cdot \prod \delta(\hat{E}_a-E_a) \times  \nonumber\\
 A_{dn}^2 (m/k)^{dn/2} (E_b-E_a)^{dn/2-1} a^{dn/2},
 \end{align}
 \begin{align}
\frac{\Omega_B^2}{E_B} = (dn/2)^2\prod \delta(\hat{E}_a-E_a) \times  \nonumber\\
 A_{dn}^2 (m/k)^{dn/2} (E_b-E_a)^{dn/2-2} E_{a}^{dn/2},
  \end{align}
 \begin{align}
\frac{\partial^2 \Phi_B}{\partial E_B^2} = dn/2\cdot (d/2-1) \times \nonumber \\
 \prod \delta(\hat{E}_a-E_a) A_{dn}^2 (m/k)^{dn/2} (E_b-E_a)^{dn/2-2} a^{dn/2}.
\end{align}
Therefore, 
\begin{align}
\frac{\partial}{\partial E_B} \biggl(T_B(E_B) \biggl) =1- \frac{\Phi_B}{\Omega_B^2} \cdot \frac{\partial^2 \Phi_B(E_B)}{\partial E_B^2} \approx 0
\end{align}
when 
\begin{align}
\frac{\partial \Phi_B}{\partial E_B} \cdot  \frac{\Phi_B}{\Omega_B^2}=\frac{\displaystyle{\frac{dn}{2}-1}}{\displaystyle{\frac{dn}{2}}} \rightarrow 1 \nonumber \\ \textrm{  as  }  dn \rightarrow \infty.  
\end{align}

When the number of particles are infinitly large, $C \rightarrow \infty$.

\section{Stochastic Thermodynamic frameworks by Jarzynski}

  We adopt the framework of Jarzynski to handle the Hamiltonian perturbation as below:
\begin{align}
H=h_s(x_j;\lambda) + H_{\mathcal{E}}(y_j) + h_{SE}(x_j,y_j) + PV_{\mathcal{E}}.
\end{align}
 The state variable $x_j$ is for the system of interest with the set of variable $\bf{Q}_j$ and $\bf{P}_j$ and then  $y_j=\left(\bf{Q}-\bf{Q}_j,\bf{P}-\bf{P}_j \right)$. $h_s(x_j;\lambda)$ is the Hamiltonian of the system of interest, which is either of $H_{\ell}$ or $H_{\theta}$. $H_{\epsilon}(y_j)$ is that of the thermal environment; therefore, it becomes the rest of the SWCNT tube system. $h_{SE}(x_j,y_j)$ is the energy related to both the system and the environment.  $P$ is the pressure of thermal environment and $V_{\mathcal{E}}$ is the volume of the total system in the reference\cite{Jarzynski2017}. The isothermal-isobaric ensemble of the system is as below: 
\begin{align}\label{eq:eq_B2}
\pi^{eq}_{\lambda NPT} (\zeta) = \frac{1}{\mathcal{Y}_A} e^{-\beta H_{S+E}(z;\lambda) + PV_{\mathcal{E}}}, \nonumber \\
 =  \frac{1}{\mathcal{Y}_A} e^{-\beta H}, \\
\mathcal{Y}_A = \int d\zeta e^{-\beta H} = \int dx e^{-\beta h_s(x;\lambda) } \mathcal{Z}^{\mathcal{E}}_{x} =\mathcal{Z}_x\mathcal{Z}^{\mathcal{E}}_0. \nonumber
\end{align}

Here, $H_{S+E}$ is $h_s(x_j;\lambda) + H_{\epsilon}(y_j) + h_{SE}(x_j,y_j) $, and $z$ is $(x_j,y_j)$. When the Eq.(\ref{eq:eq_B2}) is derived as the function of $x_j$ by the integration over $y_j$, it is the bi-Hamiltonian $h_{SE}$ that remains the perturbation to the target Hamiltonian. Therefore, the partition function of the target Hamiltonian becomes 

\begin{align}
\rho(t) = \int dy_j \pi ^{eq} (x_j,y_j),\nonumber \\ 
 = \frac{1}{\mathcal{Z_{\lambda}}}e^{-\beta (h_s(x_j;\lambda)+Pv_s(x_j)) },\\
 \mathcal{Z_{\lambda}}(N,P,T)=\int dx_j e^{-\beta(h_s+Pv_s)}.
 \end{align}
 
 $Pv_s$ is the energy that is worked to the target system from the external environment, which is equivalent to the perturbation $\phi$ from the Dzhanibekov effect in our system. 
Therefore, the probability density function for the target system is
\begin{align}\label{eq:eq_parti}
\rho = p^{eq} = \frac{e^{-\beta [h_s + \phi]}}{\int dx e^{-\beta [h_s+\phi]}} = \frac{A}{Z}.
\end{align}

\section{Derivation}

The presence of a spontaneous perturbation subject to a harmonic oscillator or Hamiltonian indicates energy loss or gain in the system. As introduced in the main text, it is so when two harmonic Hamiltonian shares the series of inertia through Dzhanibekov effect. This paper examines the possibility of using an overdamped process to resolve the conflict. Ideally, this process can substitute for the fast dynamics typically omitted when coarse-graining the full atomic system.

Later, we mention about the exponential term that appears from the overdamping process in Discussion or another publication, which allows to us to adapt the strict concept of bi-Hamiltonian and its structure from integrability and hierarchy in semi classical limit. However, for the initial set up, the notation for Hamiltonian operator in bi-Hamiltonian structure can be adapted roughly to show the effect of perturbation from Dzhanibekov effect as $\pi_{\theta}=\pi_{\theta,0}+\lambda \pi_{\ell_D}$ with $\pi_{\theta,0}$ and $\pi_{\theta}$ are from the poission tensor of the harmonic Hamiltonian for bending without and with the Dzhanibekov effect from another Hamiltonian for bond length, which affects the evolution and trajectory of evolution as noted $\pi_{\ell_D}$ with an arbitrary constant $\lambda$. About the notation, $\pi$, it means a vector filed in the trajectory of equation of motion from Hamiltonian, so that $X_H$ can be written as $X^i_H = \sum_i \pi^{ij}\frac{\partial H}{\partial x^{j}}$.  

From the previous and also naive trial of adding two independent Hamiltonian to share their momentum in CGMD simulation, we empirically know that $\pi_{\theta}$ does not satisfy the Jakobian identity, meaning that $\pi_{\theta}$ does not introduce the trajectory of Hamiltonian dynamics, and nor for vice versa for $\pi_{\ell}$. However, we can make a clear notation from borrowing the bi-Hamiltonian definition to show that the total energy of the target system, $h_s$ that follows the trajectory $\pi_{\theta}$ or $\pi_{\ell}$  is intrinsically affected by the dynamics of another Hamiltonian from Dzhanibekov effect, which we note as $u_D$ in the following paragraph, and the overdamping process from Smoluchowski operator, $L_S$ is strictly for the amount of the energy derived from $\lambda \pi_{i_D}$ with $i=\ell$ or $\theta$.  

$\rho$ is the ensemble marginalized over the state variables in $h_s$ except for $u_D$ which is from perturbation speicified as Dzhanibekov effect. $\phi$ is the probability density function, defined following Jarzynski's frameworks\cite{Jarzynski2017} as the ensemble in Eq.(\ref{eq:eq_parti}), and it is also represent the energetic term from the fast dynamics for macroscopic reaction against $u_D$, which does not exist in atomic scale dynamics. The LHS of Eq.(\ref{eqn:smolu}) divided by $\xi$ becomes

\begin{align}
\frac{\partial \rho}{\partial t} = \frac{A}{Z} \left( -\beta \frac{\partial}{\partial t}(u_D + \phi) \right).
\end{align}

The first term of RHS of Eq. (\ref{eqn:smolu}) is:
\begin{align}
\frac{\partial}{\partial x} \left(\mathcal{U}' \rho \right) = \frac{\partial ^2 \mathcal{U}}{\partial x^2} \rho + \mathcal{U}' \frac{\partial \rho}{\partial x} \\
\frac{\partial \rho}{\partial x} = - \frac{A^2}{Z^2} -\beta \frac{A}{Z} \frac{\partial}{\partial x} (u_D + \phi).
\end{align}

The second term of RHS of Eq.(\ref{eqn:smolu}) is as below:
\begin{align}
\frac{1}{\beta} \frac{\partial^2 \rho}{\partial x^2} = \frac{1}{\beta} \frac{\partial}{\partial x} \left( \frac{\partial}{\partial x} \left( \frac{A}{Z}\right) \right), \\
=\frac{1}{\beta} \left( 2\frac{A^3}{Z^3} + \frac{A^2}{Z^2} \frac{\partial ^2 Z}{\partial x^2} + \frac{1}{Z} \frac{\partial^2 A}{\partial x^2} \right).
\end{align}

With the assumption of $\frac{A}{Z}<<1$, we need the restriction on $|\frac{\partial A}{\partial x}| << 1/\beta = kT$ so that the high order of $\frac{A}{Z}$ or $|\frac{\partial A}{\partial x}|$ can be eliminated. The full derivation of Eq.(\ref{eqn:smolu}) is as below:
\begin{align}\label{eq:eq_AZ}
\xi \frac{A}{Z} \frac{\partial}{\partial t} \left(u_D + \phi \right), \nonumber \\
= \frac{\partial^2 \mathcal{U}}{\partial x^2} \frac{A}{Z}-\frac{1}{\beta} \left(-\beta \frac{A}{Z}\frac{\partial ^2}{\partial x^2} \left(u_D + \phi \right) \right).
\end{align}

For arbitrary given condition that satisfies Eq.(\ref{eq:eq_AZ}) for all $h_s$ and $\phi$ along the tube axis, Eq.(\ref{eq:eq_AZ}) becomes valid with two seperate equations which are 

\begin{align}
\frac{\partial u_D}{\partial t} =\frac{1}{\xi} \frac{\partial^2 E}{\partial x^2}, \label{eq:eq_KE} \\
\frac{\partial \phi}{\partial t}= \frac{1}{\xi}\frac{\partial^2 \phi}{\partial x^2}. \label{eq:eq_neq}
\end{align}

$E$ is the kinetic energy in $u_D$. The result of derivation indicates following points. First, any perturbation under overdamping process that is remained as an ensemble after marginalization has diffusion equation as its governing eqaution. Second, the Dzhanibekov effect that is intrinsically given from operating two Hamiltonian evolution simultaneously that causes $u_D$ can be diminished through overdamping process as written in Eq. (\ref{eq:eq_KE}). Lastly, the compensation of $u_D$ which is not included in conventional CGMD simulation, the artificially given perturbation $\phi$ should execute $u_D+\phi =0$. Since $u_D$ is on its evolution along diffusion process if it is overdamped through fast dynamcis, the perturbation from $\phi$ which is equivalent to $-u_D$ are applied through the diffusino process over the kinetic energy that can be quantified through Dzhanibekov effect as introduced in Appendix D. From this series of derivation, the compensation through $\phi$ and its double partial can be justified. 
 
 The overdamping process applied to this structure is developed through the eigenfunction $\Psi_1$ in Eq. (\ref{eq:eq_eigenfunction2}), which may allow adjustment of the Whittaker function as another form of eigenfunction for Eq. (\ref{eq:eq_KE}). This allows dynamics with a double partial derivative in the evolution process in Eq.(\ref{eq:eq_KE}). We will discuss this further in another publication. 

\section{Quantification of $\phi$}
The amount of $\phi$ is quantifiable through the trajectory of $\mathbf{Q}_{\theta}$ and $\mathbf{Q}_{\ell}$, which are the set of vectors for the coordinates of the CG particle. During the short time step, $\Delta t$, $\mathbf{Q} = \mathbf{Q}_{\ell}+\mathbf{Q}_{\theta}$ is adjusted with

\begin{align}\label{eq:eq_disp}
\Delta_t \mathbf{Q} = \left( \mathbf{P}_{\ell}+{\mathbf{P}}_{\theta}\right)\Delta t,
\end{align}

where $\mathbf{P}_{j} = P_{j}\hat{\mathbf{e}}_{j}(t)$ with $P_{j}=[P^1_{j},...,P^N_{j}]^T$ that is the set of the norm of the each momentum in $\mathbf{P}_{j} $ along the axis $\hat{\mathbf{e}}_{j}(t)$. $j$ is either $\theta$ or $\ell$.  The coordinate vector defined along the axis aligned on the new conformation at $t+\Delta t$ is 
\begin{align}\label{eq:eq_Q}
\Delta_t {Q}_{j}/\Delta t  
                            =  \left( P_{i}\hat{\mathbf{e}}_{i}(t) + P_{j}\hat{\mathbf{e}}_{j}(t) \right) \cdot\hat{\mathbf{e}}_{j}(t+\Delta t), 
\end{align}
with $j$ that is either $\theta$ or $\ell$ when $i$ is either $\ell$ or $\theta$, respectively. $\Delta_t {Q}_{j}$ is the set of the norm of each coordinate in $\mathbf{Q}_{j}$. 

From Eq.(\ref{eq:eq4}) and Eq(\ref{eq:eq5}), we can confirm that  $\Delta_t {Q}_{\theta}/\Delta t$ and $\Delta_t {Q}_{\ell}/\Delta t$ are weakly correlated each other through the Dzhanibekov effect.  
In the previous study\cite{Koh2021}, the CG model built from averaging the trajectory information from atomic simulation for cantilevered SWCNT confirms the cross correlated states between $\Delta_t {Q}_{\ell}/\Delta t$ and $\Delta_t {Q}_{\theta}/\Delta t$ in the frequency domain.

From Eq.(\ref{eq:eq_Q}),  the norm of each momentum of the CG particle becomes 
\begin{align}
{P}_j \propto \frac{\Delta_t {Q}_{j}}{\Delta t} = {P}^{H_j}+ {P}^{\Delta H_j},
\end{align}
where
\begin{align}
{P}^{H_j} = P_{j}\hat{\mathbf{e}}_{j}(t)\cdot\hat{\mathbf{e}}_{j}(t+\Delta t),\label{eq:eq_H_j}\\
{P}^{\Delta H_j} =  P_{i}\hat{\mathbf{e}}_{i}(t) \cdot\hat{\mathbf{e}}_{j}(t+\Delta t).\label{eq:eq_DH_j} 
\end{align}

 $j$ that is either $\theta$ or $\ell$ when $i$ is either $\ell$ or $\theta$, respectively. 
The perturbation energy induced by ${P}^{\Delta H_j}$ with  $j=\ell$ or $\theta$ is affected by overdamping process marked with $L_S$ in Eq.(\ref{eq:eq_op}) that causes $\Delta H$ in Eq.(\ref{eq:eq_perturb1}) and Eq.(\ref{eq:eq_perturb2}). The perturbation term in the momentum from the evolution with operator $L_S$ then shares its definition of the energy $\phi$ as    

\begin{align}
\phi \propto {P}^{H_j} \cdot {P}^{\Delta H_j},\label{eq:proj}\\ 
\Delta_t \phi = \frac{\partial^2 \phi}{\partial x^2} \Delta t. \label{eq:eq_cro_}
\end{align}

Note that  ${P}^{\Delta H_j}$ in Eq.(\ref{eq:eq_DH_j})  is from the correlated states between unit vectors that are defined from the deformation of the structure. 

The energy from $\Delta_t \phi$ is equivalent to $\Delta H$ in Eq.(\ref{eq:eq_perturb1}) and Eq.(\ref{eq:eq_perturb2}). As a result, $\Delta_t \phi$ adjusts the equation of motion in Eq.(\ref{eq:eq_go2}) with $\mathbf{P}^{\Delta H_j}$ that is governed by Eq.(\ref{eq:eq_cro}). The computation takes the form of Eq.(\ref{eq:eq_cro_}). 
 
\section{Simulation Condition}
For the MD simulation, the AiREBO potential energy function\cite{Stuart2000} is used for carbon atoms in the SWCNT. Both tubes are arranged at 300 K with 1 ns of thermalization using Langevin thermostat\cite{Schneider1978} and 1 ns of relaxation. The artificial bending is applied after the relaxation. Fixation at both ends has zero kelvin for a unit cell of the tube. Lammps packaged\cite{Thompson2022} is used. The sampling is conducted for each 1 ps during the total simulation time, which is 25 ns.  

The same arrangement is prepared using coarse grained molecular dynamics model. The heat diffusion damping, which considers the abnormal cross correlation between two independent Hamiltonians in a simple bead system, is adapted. More detail is in the reference\cite{Koh2021} and the git hub code \cite{Koh2021_}. The initial configuration of the simple bead system is derived from the molecular dynamics simulation by averaging the locations of the beads and their velocities for each unit cell, which becomes a bead.  

The sampling frequency used to count the histogram in Fig. 4 is a vital condition for the distribution plot. Without a finer sampling frequency, the CGMD simulation with HD shows very few cases that show the skewed symmetry. The total number of cases in the histogram differs between MD and CGMD cases, indicating that the event of $\phi$ for the skewed symmetry cases occurs more commonly. The sampling frequency for MD simulation is 1 ps and 0.05 ps for CGMD simulation. The couple of outliers measured in the MD simulation is discarded.    

\bibliography{SC_collision_240124_.bib}
  
\end{document}